  \documentclass{cjour}
\usepackage{graphicx}
\usepackage{amssymb}
\usepackage{amsmath}
\usepackage{algorithmic}

\newcommand{\hubc}{{\rm H}_0}

\def\sc{{\rm Science\ }}

\newcommand{\force}{{\boldsymbol{f}}}

\newcommand{\ebold}{{\boldsymbol{e}}}
\newcommand{\xbold}{{\boldsymbol{x}}}

\newcommand{\ubold}{{\boldsymbol{u}}}
\newcommand{\ibold}{{\boldsymbol{i}}}

\newcommand{\Dim}{{\mathbf{D}}}

\newcommand{\beqa}{\begin{eqnarray*}}
\newcommand{\eeqa}{\end{eqnarray*}}
\newcommand{\bgthr}{\begin{gather*}}
\newcommand{\egthr}{\end{gather*}}

\newcommand{\Domain}[1]{\ifmmode\mbox{\tt Domain<}#1\mbox{\tt>}\else{\tt Domain<$#1$>}\fi}

\newcommand{\fnc}[2]{\ifmmode\mbox{\tt#1(}#2\mbox{\tt)}\else{\tt#1(}$#2${\tt)}\fi}

\newcommand{\half}{\frac{1}{2}}

\newcommand{\IEFab}[1]{\ifmmode\mbox{\tt IEFab<}#1\mbox{\tt>}\else{\tt IEFab<$#1$>}\fi}

\newcommand{\Point}[1]{\ifmmode\mbox{\tt Point<}#1\mbox{\tt>}\else{\tt Point<$#1$>}\fi}

\newcommand{\RectDomain}[1]{\ifmmode\mbox{\tt RectDomain<}#1\mbox{\tt>}\else{\tt RectDomain<$#1$>}\fi}

\newcommand{\sign}[1]{\ifmmode\mbox{sign}(#1)\else sign(#1)\fi}
\newcommand{\Vector}[1]{\ifmmode\mbox{\tt Vector<}#1\mbox{\tt>}\else{\tt Vector<#1>}\fi}

\begin{document}









\today\\

\title{Block Structured Adaptive Mesh and Time Refinement for Hybrid, Hyperbolic + N-body Systems}


\author{Francesco Miniati$^{\dagger\ddagger}$ and Phillip Colella$^*$}
\affil{$^\dagger$Physics Department, Wolfgang-Pauli-Strasse 16, ETH Z\"urich, CH-8093 Z\"urich \\
$^\ddagger$Max-Planck-Institut f\"ur Astrophysik, 
Karl-Schwarzschild-Str. 1, 85740, Garching, Germany \\
$^*$Lawrence Berkeley National Laboratory, 1 Cyclotron Rd., 
Berkeley, CA 94720, USA
}

%









\abstract{ We present a new numerical algorithm for the solution of
  coupled collisional and collisionless systems, based on the block
  structured adaptive mesh and time refinement strategy (AMR). We
  describe the issues associated with the discretization of the system
  equations and the synchronization of the numerical solution on the
  hierarchy of grid levels.  We implement a code based on a higher
  order, conservative and directionally unsplit Godunov's method for
  hydrodynamics; a symmetric, time centered modified symplectic scheme
  for collisionless component; and a multilevel, multigrid relaxation
  algorithm for the elliptic equation coupling the two components.
  Numerical results that illustrate the accuracy of the code and the
  relative merit of various implemented schemes are also presented.}

\keywords{higher-order Godunov methods, adaptive mesh refinement, elliptic methods, particle-in-cell methods}

{{\it Mathematical subject codes:} 65M55, 76M28}

\begin{article}



\section{Introduction}

Astrophysical systems are typically complex and highly nonlinear,
providing the ground for the occurrence, either singly or 
concomitantly, of numerous physical processes. These include,
among others, hydro- and magnetohydro-dynamics, gravity, 
radiation and many-body interactions.
They operate on a wide range of spatial and temporal scales and it
is often desirable to fully cover these ranges for a thorough
understanding of the problem. 
Thus the problems are demanding both in  terms
of physics algorithms and dynamic range (resolution). 
While the development of high order numerical schemes certainly 
improves the quality of numerical solutions and the availability of 
ever more powerful computers has allowed the performance of larger
calculations, special techniques are required in order to achieve
very large dynamic ranges. 

The Adaptive Mesh Refinement (AMR) technique offers a powerful
solution for this purpose~\cite{bergeroliger84,bergercolella89}.
While there are difficulties associated with its implementation,
the use of AMR in astrophysics and cosmology has grown significantly
to include studies of nucleosynthesis in Supernovae 
explosions~\cite{timmesetal00,kifonidisetal00,almgrenETAL:2006a,almgrenETAL:2006b}, 
of the multiphase interstellar medium and radiative 
shock hydrodynamics~\cite{kmc94,wafo98}, 
the problem of star formation out of the collapse of protostellar
clouds~\cite{kleinetal03,trueloveetal97,trueloveetal98} and
the formation of the first stars as well as the large scale structure 
in the universe~\cite{abn02,norman04b}.
The use of AMR technique in the above examples has often been 
instrumental in either revealing new properties of the investigated 
system (such as instabilities) or pointing out mistaken views based
on limited resolution calculations.

We will consider two aspects of application of AMR to astrophysical
problems. The first is the extension to incorporate self-gravity.
This capability leads to new algorithmic difficulties due to the
elliptic nature of the problem.  In particular the solution has to be
computed simultaneously on all levels of refinement and continuity of
both the solution and its normal derivative has to be enforced at
coarse-fine level interfaces ~\cite{maca96}.  However, this introduces
non-trivial complications when time refinement is also employed: the
coarser levels are advanced first and finer levels are advanced with
the assumption that boundary conditions at fine-coarse interfaces are
provided by the coarse level solutions interpolated in time. However,
the full multilevel elliptic solution can only be computed when all
levels are synchronized and thus is not available when the coarser
levels are ahead of the finer levels.  Thus the first implementations
of a full multilevel elliptic solver for
self-gravity~\cite{trueloveetal97,trueloveetal98,rickeretal05} do not
use refinement in time. And when employing time refinement the
multilevel elliptic equation has been solved as a set of independent
boundary value problems, one for each level, not a fully multilevel
solution~\cite{kravtsov99,knebeetal01,teyssier02,quilis04}.

The second issue we will address is the application of AMR to hybrid
systems, that is a self-gravitating gas coupled to a particle
representation of a collisionless matter described by Vlasov-Poisson
equations~\cite{hoea81} .  This is relevant to several problems in
astrophysics, particularly for modeling the formation and evolution of
structure in the universe.  In this case the AMR technique is combined
with Particle-Mesh methods to compute the right-hand side to the
Poisson's equation due to the particles mass.  In order to take
advantage of the higher resolution of the finer grids, it is desirable
to advance the particles with the force compute on, and the timesteps
of, the finest grids that cover their spatial position. This
introduces a complication for the elliptic solver similar to the one
described above for the case of self-gravitating gas dynamics, in the
sense that the particles contribution to the right-hand side of
Poisson's equation needs to be accounted for even when the levels of
refinement are not synchronized.

Various AMR codes for hybrid systems have been developed over the
past several
years~\cite{bryannorman97,kravtsov99,knebeetal01,teyssier02,quilis04,rickeretal05}.
The refinement strategy in~\cite{kravtsov99,teyssier02,rickeretal05}
is based on splitting of individual cells and in~\cite{knebeetal01}
only the collisionless component is evolved.  Virtually all schemes
use Strang splitting~\cite{strang68} for the multidimensional version
of the hydrodynamics and a modified leap-frog method for the
integration of the equation of motion of the particles.  This has also
been done for problems in which there is no coupling to a collisional
fluid component, as occurs in computations of collisionless 
plasmas~\cite{vayETAL:2002}.

In our approach we use the block structured scheme for adaptive mesh
and time refinement proposed in~\cite{bergercolella89} as a starting
point and extend it include gravity and collisionless particle
dynamics.  We use an unsplit Godunov's method for hydrodynamics
\cite{millercolella02}; a symmetric, time centered modified symplectic
scheme based on the kick-drift-kick sequence for the collisionless
component; and a multilevel, multigrid relaxation algorithm for the
elliptic equation coupling the two systems.  We introduce two new
procedures to solve synchronization issues described above that arise
with the elliptic solver when the coarse and finer levels are not
synchronized. We use a method analogous to those developed for AMR for
incompressible flows \cite{almgrenETAL:1998,martinColella:2000} to
compute a lagged estimate of the correction of the elliptic matching
conditions at boundaries between refinement levels at times when the
levels are not synchronized.  We also present a detailed discussion of
refinement in time in the presence of collisionless particles,
including methods for associating particles with refinement levels,
and a particle aggregation operation to cost-effectively estimate the
density distribution on coarse levels due to particles evolved on
finer levels without compromising the code accuracy. We provide a
detailed description of the formal discretization of the system of
equations and the issues involved with the synchronization of the
numerical solution in the presence of refinement in time, at a level
of detail which we feel is lacking in the current literature.

The paper is organized as follows.  First, the evolution equations
and single level algorithms are outlined in Section~\ref{sleq.sec}. In
Section~\ref{amrgf.sec} we provide a formal definition of the
employed AMR volume discretization, variables and operators and then
describe in detail the general AMR algorithm for hybrid system.
Finally, in section~\ref{tests.sec} we test the accuracy of the code
by comparing its results against for set of standard solutions.

\section{Evolution Equations and Temporal Discretization}
\label{sleq.sec}

In this section we introduce the system equations and describe their
temporal discretization on individual levels of refinement. Motivated
by {\it cosmological} applications, the numerical schemes are
formulated for a grid with a time dependent scale length, $a(t)$.
Thus after a description of the expanding grid, we introduce the
time discretization for the equations of hydrodynamics with 
gravity and the equations of motion for the collisionless component
and then briefly outline the time step constraints and code units.
\subsection{Comoving Frame}
Cosmic expansion is described by the first Friedmann's equation which
reads~\cite{peebles93}
\begin{equation} \label{einstein1.eq}
\frac{\dot{a}}{a} =  \hubc \; 
\left(\Omega_m \, a^{-3} + \Omega_k \, a^{-2} + \Omega_\Lambda \right)^{1/2}
\end{equation}
where $a(t)$ is the scale of the universe as a function of
time; $H_0$ measures the current ($a=1$) rate of cosmic expansion (the
Hubble constant); $\Omega_m, \Omega_k$ and $\Omega_\Lambda$ are
parameters representing the current energy density associated to matter,
curvature and `dark' component, respectively, in units of the closure value.
The solution to Eq.~(\ref{einstein1.eq}) relating cosmic time and expansion 
parameter reads
\begin{equation} \label{tepx.eq}
H_0\, t(a) = \int_0^a \frac{d\tilde{a}}{\tilde{a}\,
\left(\Omega_m \, \tilde{a}^{-3} + \Omega_k \, \tilde{a}^{-2} + \Omega_\Lambda \right)^{1/2} }
\end{equation}
and admits simple solutions for $\Omega_k=0$. 
Since the equations of motion in an expanding Universe are most naturally
solved in a {\it comoving} frame, which expands at the rate
$\dot{a}/a$ given by Eq. (\ref{einstein1.eq}), we operate the
change of coordinates
\begin{equation}\label{comfr.eq}
{\boldsymbol x} = a(t)^{-1}\;{\boldsymbol r} 
\end{equation}
where ${\boldsymbol r}$ and ${\boldsymbol x}$ are the coordinates in
the laboratory and comoving reference frame respectively, and transform
all differential operators (time derivatives, gradient, laplacian) 
according to~\cite{peebles93}
\begin{eqnarray}  \label{dertr1.eq}
\frac{\partial }{\partial{\boldsymbol r}} & \rightarrow  & \frac{1}{a}
\;\frac{\partial}{{\boldsymbol x}} \\ \label{dertr2.eq}
\left(\frac{\partial}{\partial t}\right)_{\boldsymbol r} & \rightarrow  &
\left(\frac{\partial }{\partial t}\right)_{\boldsymbol x} +
\frac{\partial {\boldsymbol x}}{\partial t}\;
\left(\frac{\partial }{\partial{\boldsymbol x}}\right)_t
=
\left(\frac{\partial }{\partial t}\right)_{\boldsymbol x} - \frac{\dot{a}{\boldsymbol x}}{a}\left(\frac{\partial }{\partial{\boldsymbol x}}\right)_t .
\end{eqnarray}
The velocity 
\begin{equation}
\dot{\boldsymbol r}=\dot{a}{\boldsymbol x}+a\dot{\boldsymbol x}\equiv\frac{\dot{a}}{a} 
{\boldsymbol r}+a\dot{\boldsymbol x}
\label{cotra.eq}
\end{equation}
is decomposed into a Hubble flow, $ (\dot{a}/a)\,
{\boldsymbol r}$, and a peculiar proper component, $u = a\,
\dot{\boldsymbol x}$. It is also convenient to introduce the density
and pressure in terms of the comoving volume $x^3$, as opposed to the
proper volume $r^3$, 
\begin{eqnarray}  
\label{procom1.eq}
\rho(t,{\boldsymbol x}={\boldsymbol r}/a) & = & a^3 \; \rho_p (t,{\boldsymbol r})\\
P (t,{\boldsymbol x}={\boldsymbol r}/a)  & = & a^3 \; P_p(t,{\boldsymbol r})  
\label{procom2.eq}
\end{eqnarray}
where the subscript $p$ indicates the proper quantities.
\subsection{Hydrodynamics} \label{hydro.sec}
In comoving coordinates, the hydrodynamics is described by the
following set of inhomogeneous partial differential equations
\begin{eqnarray}
\frac{\partial \rho}{\partial t} + \frac{1}{a}
\frac{\partial}{\partial x_k} (\rho u_k) & = & ~~ 0  \label{rhoe.eq}\\
\frac{\partial \rho u_i}{\partial t} + \frac{1}{a} 
\frac{\partial}{\partial x_k} ( \rho u_i u_k +P\, \delta{ik})
 & = &  -\frac{\dot{a}}{a}\,\rho u_i - \frac{1}{a}\,\rho 
\frac{\partial \phi}{\partial x_i}   \label{mome.eq} \\
\frac{\partial \rho e}{\partial t} + \frac{1}{a}
\frac{\partial}{\partial x_k} [(\rho e+P) u_k] 
 & =  & -2\frac{\dot{a}}{a}\,\rho e - \frac{1}{a}\,\rho u_k 
\frac{\partial \phi}{\partial x_k}    \label{enee.eq}\\
\frac{\partial \rho s}{\partial t} + \frac{1}{a} 
\frac{\partial}{\partial x_k} (\rho s u_k) & = & -2\frac{\dot{a}}{a}\,\rho s 
\label{entre.eq}
\end{eqnarray}
expressing (from top to bottom) mass, momentum, energy and entropy
conservation. Here, $\rho$ and $P$ are the comoving density and
pressure respectively, $u$ is the peculiar proper velocity, $\phi$ the
proper gravitational potential, $e\equiv e_{th} + e_k = P/ \rho
(\gamma-1) + u^2/2 $ is the specific total energy, $s=P/\rho^\gamma$
is the specific entropy and $\gamma$ the gas adiabatic index.  The
first inhomogeneous terms on the right hand side of Eq.
(\ref{mome.eq})-(\ref{entre.eq}) describe the effects due to adiabatic
cosmic expansion ($\propto \dot{a}/a$). In particular, the factor 2
for the last two equations arises by assuming that the internal energy
of the gas is solely associated with translational degrees of
freedom\footnote{More generally, in $D$ dimensions, the energy losses due to expansion are $\dot{e}_{th}/e_{th} = D(\gamma-1) \dot{a}/a$ for the internal energy and $\dot{e}_{k}/e_{k} = 2\dot{a}/a$ for the specific kinetic energy respectively.}.  Finally, gravity is
described by the gravitational potential $\phi$, which is generated by
the matter distribution of both the collisional and the collisionless
components.  

We use a cell-centered discretization for our primary dependent
variables: $U(\xbold,t) \equiv (\rho,\rho {\bf u},\rho e)^T
\rightarrow U^n_\ibold$, where $\ibold \in \mathbb{Z}^\Dim$ indexes
grid points in a space with $\Dim$ dimensions and $n$ is
a discrete time index.  The starting point for our temporal
discretization is a conservative finite-difference method for the
hydrodynamic equations:
\begin{eqnarray}\label{usplit.eq}
U^{n+1,h} & = & U^n -  \Delta t \,{\cal A}(t)
\, \left(D \cdot \vec{F}^{n + \half}\right)
\end{eqnarray}
where 
$D \cdot \vec{F}^{n + \half}$ approximates the spatial
derivative terms on the left-hand side of (\ref{enee.eq}), 
${\cal A}(t)$ is a diagonal matrix with elements
$\left[\frac{1}{a^{n+1/2}},\frac{1}{a^{n+1}},
\frac{a^{n+1/2}}{(a^{n+1})^2},\frac{a^{n+1/2}}{(a^{n+1})^2}\right]$,
and $\Delta
t$ is the time step. We use an unsplit second-order Godunov's method
\cite{colella90,saltzman94,millercolella02} to compute the flux
divergence.
In addition to the cosmological expansion terms, the gas and particle 
components couple through the force field which
is solution to the following Poisson's equation
\begin{eqnarray} \label{poisson.eq}
\Delta \phi & = & \frac{4\pi G}{a}\,\left(\rho_m-\langle\rho_m \rangle\right).
\end{eqnarray}
Here $\rho_{\rm m} = \rho_{\rm gas} + \rho_{\rm part}$ is the total
comoving mass density; the particle density, $\rho_{\rm part}$, is
computed through a Particle Mesh method (see details in the
Appendix). When periodic boundary conditions are used,
$\langle\rho_m\rangle$ is the volume average, otherwise it is zero.
The details of the spatial discretization for Poisson's equation are
given in the next section. For now, we assume that we can compute the
gravitational acceleration at cell centers, $\force_{\ibold} \approx
-\nabla \phi$, with second order accuracy.

To compute the effect of the source terms, we need to compute them
before and after the hydrodynamic update, taking advantage of the fact
that no sources appear in the density evolution equation:
\begin{equation}
S_{{\boldsymbol i}}^n = 
 \left( \begin{array}{c} 
0 \\ 
 \rho_{\boldsymbol i}^{n+1/2} \force^n_\ibold /a^{n+1} + 
\rho_{\boldsymbol i}^n {\boldsymbol u^n_\ibold} \, [(a^n/a^{n+1})-1] \\ 
\Delta (\rho e_k) + P_{\boldsymbol i}^n \, [(a^n / a^{n+1} )^2 -1]  \\ 
\rho_{\boldsymbol i}^n s_{\boldsymbol i}^n \, [(a^n / a^{n+1} )^2 -1] 
\end{array} \right) 
\label{eqn:source}
\end{equation}
where $\rho_{\boldsymbol i}^{n+1/2} = (1/2)\,(\rho_{\boldsymbol i}^{n}+\rho_{\boldsymbol i}^{n+1})$, and $\Delta (\rho e_k)=(1/2)\,[\rho_{\boldsymbol i}^{n+1} (u_{\boldsymbol i}^{n+1})^2 - \rho_{\boldsymbol i}^{n} (u_{\boldsymbol i}^{n})^2]$.
After the hydrodynamics update given in Eq.(\ref{usplit.eq})
we apply the source terms as follows
\begin{equation} 
U^{n+1,h,s_1}   =  U^{n+1,h} + \Delta t \, S^n.
\end{equation}
The above estimate of the source term, 
after being converted to primitive variable form,
is also used in the hydrodynamic
predictor step in order to obtain fully time-centered fluxes.
$S^n$ accounts for the expansion terms to the desired accuracy but is
only first order accurate as far as the gravity term is concerned
(hence the superscript $s_1$).  After the new gravitational potential
has been computed a
source correction term is estimated as
\begin{equation}
\delta S^{n+\frac{1}{2}}_{{\boldsymbol i}} =
\frac{1}{2} \rho^{n+1}_{\boldsymbol i}\,\frac{\delta\force_\ibold^{n+1/2}}{a^{n+1}}
 \left( \begin{array}{c} 
0 \\ 
1 \\
{\boldsymbol u}_\ibold^{n+1}+\left[\Delta t/4\,a^{n+1}\right]\,\delta\force_\ibold^{n+1/2} \\ 
0 
 \end{array} \right) 
 \label{eqn:forceUpdate}
\end{equation}
with, $\delta\force^{n+1/1}_\ibold=\force_\ibold^{n+1}-\force_\ibold^n$,
and the final source update is
\begin{equation} 
U^{n+1} =   U^{n+1,h,s_1} + \Delta t \, \delta S^{n+\frac{1}{2}}.
\label{eqn:ds}
\end{equation}
\subsubsection{Hypersonic Flows}
Accretion flows induced by gravity are typically hypersonic and can
be characterized by very large Mach numbers ${\cal M} \geq 100$. This
situation is common in cosmological simulations~\cite{minetal00}.

In this case the total energy is largely dominated by the kinetic
component, $e_{k}\sim {\cal M}^2\,e_{th}$. Since conservative
hydro-schemes track the total energy, $e= e_{th}+e_{k}$, relatively
small errors in the partition of the two components can produce
spurious values of $e_{th}$. This is particularly worrisome when a
4-byte (single precision) digital representation of the numerical data
is employed. For this reason, we introduce the additional equation
(\ref{entre.eq}) describing the evolution of the gas entropy.  When
the Mach number of the bulk flow is very high, ${\cal M}\ge 50$,
and away from shocks, it
provides a more accurate solution from the thermal energy than the
total energy equation. Eq. (\ref{entre.eq}) is naturally incorporated
in the numerical scheme for hydrodynamics adopted here because of its
conservative form.  In addition, being a simple advection equation,
the conservative fluxes for its integration are a byproduct of the
Riemann solution and require virtually no extra effort to compute.

The equation for the internal energy could be alternatively integrated
~\cite{bryanetal95}, but its non-conservative form makes it less
attractive. Authors in Ref.~\cite{rokc93} already employed the entropy
equation in order to improve the accuracy of their Total Variation
Diminishing scheme for hypersonic flows, although their implementation
required solving twice for the hydrodynamic equations with an extra
cost of 30\%. 

It is worth pointing out that for high Mach number flows, errors
in the velocity field are propagated into the Riemann solver solution for the
pressure at the cell interface, $P^*$, with a coefficient $\sim {\cal
M}$, that is, largely amplified. More precisely we find
\begin{equation}
\frac{\delta P^*}{P^*} = \gamma \frac{\delta \Delta u}{u} {\cal M}
\end{equation}
where $\Delta u$ is the one dimensional velocity jump at the cell interface.
However, as illustrated above, the pressure terms enter the
hydrodynamic equations with a weight ${\cal M}^{-1}$ as compared to
kinetic terms and, therefore, the numerical solution is not degraded
in the case of high Mach number flows because of approximations
involved in the Riemann solver solution.

\subsection{Collisionless Component} \label{particle.sec}
The collisionless component is described by a set of particles whose
evolution in phase space is computed according to 
\begin{eqnarray}
\frac{d {\boldsymbol y}}{dt} & = &~~\frac{1}{a}\, {\boldsymbol u} \label{posdm.eq} \\
\frac{d{\boldsymbol u}}{dt} & = & -\frac{\dot{a}}{a}\,{\boldsymbol u} +\,\frac{1}{a}\, 
\force \label{veldm.eq}
\end{eqnarray}
where ${\boldsymbol y}$ and ${\boldsymbol u}$ are the comoving
coordinate and peculiar proper velocity respectively.  The acceleration
acting on the particle, $\force$, is obtained by
first computing the acceleration on the grid, using a cell- or face-
centered scheme, and then by interpolating it to the particle position
through a Particle Mesh method \cite{hoea81}.

In order to advance in time the particle positions and velocities
we propose the following integration scheme based on a 
kick-drift-kick sequence~\cite{quinnetal97}:
first the particles velocity and positions are updated as
\begin{eqnarray}
{\boldsymbol u}^{n+1/2}  & = &  {\boldsymbol u}^n \,  \frac{a^n}{a^{n+1/2}}
+ \frac{1}{a^{n+1/2}}\,\force^n ({\boldsymbol y}^{n})  \frac{\Delta t}{2}  
\label{eqn:particleVelocityPredictor}
\\
\label{eqn:particlePositionPredictor}
{\boldsymbol y}^{n+1}  & = & {\boldsymbol y}^{n}
+ \frac{1}{a^{n+1/2}}\,{ \boldsymbol u}^{n+1/2} \, \Delta t.
\end{eqnarray}
After computing the acceleration at the new timestep,
the particle velocity is finally updated as
\begin{eqnarray} \label{kdku.eq}
{\boldsymbol u}^{n+1}  =  {\boldsymbol u}^{n+1/2} \,  \frac{a^{n+1/2}}{a^{n+1}}
+ \frac{1}{a^{n+1}}\, \force^{n+1}({\boldsymbol y}^{n+1})  \frac{\Delta t}{2}. 
\end{eqnarray}

The proposed scheme is {\it reflexive} and hence
symplectic~\cite{KahanLi97}. This has the nice property that the
integral of motion will be conserved on average preventing secular
accumulation of error and keeping the system about its true trajectory
in phase space~(see, e.g., discussion in \cite{quinnetal97}).
We have also implemented an alternative method, 
based on the more common drift-kick-drift sequence,
which does the following:
first particle positions at half the time step are predicted as
\begin{equation} \label{predx.eq}
{ \boldsymbol y}^{n+1/2}   =  {\boldsymbol y}^n +
\frac{1}{a^{n}}\,{\boldsymbol u}^n \frac{\Delta t}{2},
\end{equation}
then particle positions and velocity are further temporarily updated to
\begin{eqnarray} \label{corru.eq}
{\boldsymbol u}^{n+1,*}  & = &  {\boldsymbol u}^n \,  \frac{a^n}{a^{n+1}}
+ \frac{1}{a^{n+1}}\, \force^n ({\boldsymbol y}^{n+1/2})\Delta t  \\
\label{corrx.eq}
{\boldsymbol y}^{n+1,*}  & = & {\boldsymbol y}^{n+1/2}
+ \frac{1}{a^{n+1}}\,{ \boldsymbol u}^{n+1,*} \, \frac{\Delta t}{2}.
\end{eqnarray}
Based on the gravitational potential at the new time-step, a final
correction term is applied, that allows second order accuracy
\begin{eqnarray} \label{corru2.eq}
{ \boldsymbol u}^{n+1}  & = &  {\boldsymbol u}^{n+1,*}
+ \frac{1}{a^{n+1}}\, 
\left[ \force^{n+1} ({\boldsymbol y}^{n+1/2}) - \force^n ({\boldsymbol y}^{n+1/2})\right]
\frac{\Delta t}{2} \\
\label{corrx2.eq}
{\boldsymbol y}^{n+1}  & = &  {\boldsymbol y}^{n+1/2} 
+ \frac{1}{a^{n+1}}\,{\boldsymbol u}^{n+1} \frac{\Delta t}{2}.
\end{eqnarray}
The above scheme, however, is not fully reflexive.  The need of
temporary states that approximate the solution at the end of the
time step in order to calculate the final correction step, breaks the
time symmetry of the scheme.

Note that in both schemes, there is no need for storage of extra
information about either the particle positions or their velocities at
any old or intermediate step. In addition, the gravitational potential is
computed only once per time step making their overall
computational cost of rather inexpensive.

It should be noticed, however, that in the case of AMR reflexivity is
lost even in the former scheme without additional precautions.  This
occurs when a particle is transferred to a different level of
refinement, or even during a refinement operation. In both cases, the
gravitational potential changes as a result of these
procedures.  Since this change takes place after the correction steps,
the backward application of the scheme would not reproduce the initial
configuration.  Although in principle this could be fixed by storing
information about the last timestep and acceleration for each
particle, we have not implemented any of this.

\subsection{Time Step} \label{timestep.sec}
The time-step is subjected to the following constraints: In accord with
the Courant-Friedrichs-Lewy (CFL) condition for stability of finite
difference methods~\cite{leveque92}, we require
\begin{equation} \label{dthydr.eq}
\Delta t = C_{\rm hydro} 
\frac{a(t)\,\Delta x}{{\rm Max}(|u_i|+c_s)},
\end{equation}
where $C_{\rm hydro} <1$ is the CFL number, and $u_i$ and $c_s$ are the
fluid velocity in the $i_{th}$ direction and sound speed of the flow
respectively. 
In presence of a source term, $S$, we modify the estimate of the 
fluid velocity according to
\begin{equation} \label{usrce.eq}
|u_i|+c_s \longrightarrow \frac{|S_i|\, \Delta x}{[(|u_i|+c_s)^2 + 2\,|S_i|\, \Delta x]^{1/2}-(|u_i|+c_s)}
\end{equation}
where $S_i$ is the component of the source term affecting the velocity $u_i$.
For the purpose of accuracy, rather than stability, for the collisionless 
particles we likewise require
\begin{equation} \label{dtpart.eq}
\Delta t = C_{\rm part} 
\frac{a(t)\,\Delta x}{{\rm Max}(|u_i|)}
\end{equation}
with $C_{\rm part}<1$ and with $U_i$ corrected as in Eq.~(\ref{usrce.eq})
but with $c_s=0$.
Finally, we require that the background expansion 
remains limited during each integration cycle. This allows us
to time center the value of $a(t)$ in our integration schemes
above and neglect its changes with time. We thus enforce
\begin{equation} \label{dtexp.eq}
\Delta t < C_{\rm exp} \;\frac{a}{\dot{a}}
\end{equation}
with $ C_{\rm exp} \simeq (1-2) \times 10^{-2}$.

\subsection{Code Units}
The natural choice for the dimensional units of the above
physical equations is given by the following lengths, mass 
and time scales
\begin{eqnarray}
L_*    & = & L_{\rm box} \\
\rho_* & = & \rho_c \,\Omega_m \\
t_*    & = & H_0^{-1}
\end{eqnarray}
where$L_{\rm box}$ is the size of the computational box and $\rho_c
\equiv 3 \, H_0^2 /8\pi \,G =1.879\times 10^{-29}\; h^2\;$ is the
critical density of the universe,
with $h \equiv H_0/100$ km s$^{-1}$Mpc$^{-1}$. 
The units for the other quantities are defined in terms of these as
\begin{eqnarray}
u_*   & = &  H_0^{-1} \, L_{\rm box}     \\  
P_*   & = &  \rho_* \, u_*^2            \\
\phi_* & = & u_*^2                      \\
T_*   & = &  m_{proton} \,P_*/ k_B \rho_* 
\end{eqnarray}
where $m_{proton}$ is the proton mass and $k_B$ Boltzmann's constant.
\section{Adaptive Mesh Refinement Approach} \label{amrgf.sec}

In this section we outline the structure of the AMR algorithm.
After introducing the formal notation, we describe in some
detail the scheme for hydrodynamics with self-gravity and its 
modifications when a collisionless component is also included,
in the simple case of two levels of refinement.
We then describe the extension of the scheme to the general multilevel case.

\subsection{Multilevel Volume Discretization, Variables and Operators}

The underlying discretization of the $\Dim$-dimensional space is given
as points $(i_0, ... , i_{\Dim - 1}) = \ibold \in \mathbb{Z}^\Dim$.
The problem domain is discretized using a grid $\Gamma \subset
{\mathbb{Z}}^\Dim$ that is a bounded subset of the lattice.  $\Gamma$
is used to represent a finite-volume discretization of the continuous
spatial domain into a collection of control volumes: ${\ibold} \in
\Gamma$ represents a region of space,
\begin{equation}
V_{\ibold} = [\ibold h, (\ibold + \ubold)h]
\end{equation}
where $h$ is the mesh spacing, and $\ubold \in
{\mathbb{Z}}^\Dim$ is the vector whose components are all equal to one.
We can also define face-centered 
discretizations of space based on those control volumes:
$\Gamma^{\ebold^d} = \{\ibold \pm \half \ebold^d: \ibold \in \Gamma\}$,
where $\ebold^d$ is the unit vector in the $d$ direction.
$\Gamma^{\ebold^d}$ is the discrete set that indexes the faces of the
cells in $\Gamma$ whose normals are $\ebold^d$:

\begin{equation}
A_{\ibold + \half \ebold^d} =
 [(\ibold + \ebold^d )h, (\ibold + \ubold)h],
\ibold + \half \ebold^d \in \Gamma^{\ebold^d}.
\end{equation}
We define cell-centered discrete variables on $\Gamma$:
\begin{gather*}
\phi : \Gamma \rightarrow {\mathbb{R}}^m
\end{gather*}
and denote by $\phi_\ibold
\in {\mathbb{R}}^m$ the value of $\phi$ at cell $\ibold \in
\Gamma$.  We can also define face-centered vector fields on $\Gamma$:
\begin{gather*}
\vec{F} = (F_0, ..., F_{\Dim-1}) \hbox{ , } F_d : \Gamma^{\ebold^d}
\rightarrow {\mathbb{R}}^m
\end{gather*}
and define a discretized divergence operator on such a vector field:
\begin{equation} 
\label{eqn:flux1}
(D\cdot \vec{F})_\ibold = \frac{1}{h} \sum^{\Dim-1}_{d=0} (F_{d,\ibold
+\half \ebold^d} - F_{d,\ibold - \half \ebold^d}), \ibold \in
\Gamma.
\end{equation}
We will find it useful to define a number of operators on points and
subsets of $\mathbb{Z}^\Dim$. 
We define a coarsening operator by:
${\mathcal{C}}_r : {\mathbb{Z}}^\Dim \rightarrow {\mathbb{Z}}^\Dim$,
\begin{gather*}
{\mathcal{C}}_r(\ibold) = \left(\left\lfloor \frac{i_0}{r} \right\rfloor,...,
\left\lfloor \frac{i_{d-1}}{r} \right\rfloor \right)
\end{gather*}
where r is a positive integer.
These operators acting on subsets of $\mathbb{Z}^\Dim$ can be
extended in a natural way to the face-centered sets:
${\mathcal{C}}_r(\Gamma^{\ebold^d}) \equiv
({\mathcal{C}}_r(\Gamma))^{\ebold^d}$.
For any set $\Upsilon \subseteq \Gamma$, we define 
$\mathcal{G}(\Upsilon,r)$, $r > 0$, to be the set of all points
within a $|\cdot|$-distance $r$ of $\Upsilon$ that are still contained
in $\Gamma$:
\begin{gather*}
\mathcal{G}(\Upsilon,r) = \Gamma
\cap \underset{|\ibold| \leq r}{\cup} \Upsilon + \ibold
\end{gather*}
where $|\ibold| = \underset{d = 0 \dots \Dim - 1}{max}(|i_d|)$.
We can extend the definition to the case $r < 0$ :
\begin{gather*}
\mathcal{G}(\Upsilon,r) = \Gamma - \mathcal{G}(\Gamma - \Upsilon,-r).
\end{gather*}
Thus $\mathcal{G}(\Upsilon,r)$ consists of all of the points in $\Upsilon$
that are within a distance $-r$ from points in the complement of
$\Upsilon$ in $\Gamma$.
In the case that there are periodic boundary conditions in one or more of 
the coordinate directions, we think of the various sets appearing here
and in what follows as consisting of the set combined with all of its
periodic images for the purpose of defining set operations and computing 
ghost cell values. For
example, $\mathcal{G}(\Upsilon,r)$ is obtained by growing the union of 
$\Upsilon$ with its periodic images, and performing the intersections 
and differences with the union of $\Gamma$ with its periodic images.

We use a finite-volume discretization of space to represent a nested hierarchy 
of grids that discretize the same continuous spatial domain.
We assume that our problem domain can be discretized by a nested
hierarchy of grids $\Gamma^0 ... \Gamma^{lmax}$, with $\Gamma^{l+1} = 
{\mathcal{C}}^{-1}_{n^l_{ref}}(\Gamma^l)$, 
and that the mesh spacings $h^l$ associated with $\Gamma^l$ 
satisfy $\frac{h^{l}}{h^{l+1}} = n^l_{ref}$. 
The integer $n^l_{ref}$ is
the {\it refinement ratio} between level $l$ and $l+1$.  
These conditions imply that the
underlying continuous spatial domains defined by the control volumes 
are all identical.
In this paper we will further assume $n^l_{ref}$ is even. In the case where there are only two levels, we will refer to them as {\it coarse}
and {\it fine}, with the notation $\{l=0,l=1\} \rightarrow \{c,f\}$, and 
$n_{ref}^0 \rightarrow n_{ref}$.

We make two assumptions about the nesting of grids at successive levels.
We require the control volume
corresponding to a cell in $\Omega^{l-1}$ is either completely
contained in the control volumes defined by $\Omega^l$ or its
intersection has zero volume. We also assume that there is at least
one layer of level--$l$ cells separating level--$(l+1)$ cells from 
level--$(l-1)$ cells:
$\mathcal{G}(\mathcal{C}_{n^l_{nref}}(\Omega^{l+1}),1)
\subseteq
\Omega^l$. We refer to grid hierarchies that meet 
these two conditions as being
{\it properly nested}. 



From a formal numerical analysis standpoint, a solution on an adaptive
mesh hierarchy $\{\Omega^l \}^{l_{max}}_{l=0}$ 
approximates the exact solution to the Partial Differential Equations
only on those cells
that are not covered by a grid at a finer level.  We define the valid
region of $\Omega^l$ as
\begin{gather*}
\Omega^l_{valid} = \Omega^l - {\mathcal{C}}_{n^l_{ref}}(\Omega^{l+1}).
\end{gather*}
A composite array $\psi^{comp}$ is a collection of discrete values
defined on the valid regions at each of the levels of refinement:
\begin{gather*}
\psi^{comp} = \{\psi^{l, comp} \}^{l_{max}}_{l=0}
\mbox{  ,   }
\psi^{l, comp} : \Omega^l_{valid} \rightarrow {\mathbb{R}}^m.
\end{gather*}
We can also define valid regions and composite arrays for face-centered
variables: $\Omega^{l, \ebold^d}_{valid} = \Omega^{l, \ebold^d} -
{\mathcal{C}}_{n^l_{ref}}(\Omega^{l+1, \ebold^d})$.  Thus, $\Omega^{l,
\ebold^d}_{valid}$ consists of $d$-faces that are not covered by the
$d$-faces at the next finer level.  A composite vector field
$\vec{F}^{comp} = \{\vec{F}^{l, valid} \}^{l_{max}}_{l=0}$ is defined
as follows:
\begin{gather*}
\vec{F}^{l, comp} = (F^{l, comp}_0 \dots F^{l, comp}_{\Dim-1})
\hbox{ , } F^{l, comp}_d : \Omega^{l, \ebold^d}_{valid} \rightarrow
{\mathbb{R}}
\end{gather*}
Thus a composite vector field has values at level $l$ on all of the
faces not covered by faces at the next finer level.

We want to define a composite divergence $D^{comp}(\vec{F}^{l+1,comp},
\vec{F}^{l,comp})_\ibold$ for $\ibold \in \Omega^l_{valid}$.  To do
this, we construct an extension of $\vec{F}^{l, comp}$ to the edges
adjacent to $\Omega^l_{valid}$ that are covered by fine level faces.
On the valid coarse-level $d$-faces, $F^l_d = F^{l, comp}_d$.  On the
faces adjacent to cells in $\Omega^l_{valid}$, but not in $\Omega^{l,
  \ebold^d}_{valid}$, we set $F^l_d = \langle F^{l+1,comp}_d\rangle $,
the average of $F^{l+1}_d$ onto the next coarser level:
\begin{gather*}
\langle {F}^{l+1}_d\rangle _{\ibold_l + \half \ebold^d} = \frac{1}{(n_{ref})^{\Dim-1}}
\sum_{\ibold + \half \ebold^d \in {\mathcal{F}}^d} F^{l+1}_{d,\ibold
+ \half \ebold^d}
\hbox{ , }
\ibold_l + \half \ebold^d \in {\zeta}^{l+1}_{d,+} \cup {\zeta}^{l+1}_{d,-}.
\end{gather*}
Here ${\mathcal{F}}^d$ is the set of all fine level $d$-faces that are
covered by $A_{\ibold_l + \half \ebold^d}$.
${\zeta}^{l+1}_{d,\pm}$ consists of all the $d$-faces in $\Omega^l$ on
the boundary of $\Omega^{l+1}$, with valid cells on the low $(\pm = -)$
or high $(\pm = +)$ side:
\begin{gather*}
{\zeta}^{l+1}_{d,\pm} = \{\ibold \pm \half \ebold^d : \ibold \pm 
\ebold^d \in \Omega^l_{valid}, \ibold \in
{\mathcal{C}}_{n^l_{ref}}(\Omega^{l+1}) \}.
\end{gather*}
Given that extension, our composite divergence is defined as:
\begin{gather}
\label{eqn:compositeDiv}
D^{comp}(\vec{F}^{l+1,comp},\vec{F}^{l,comp})_\ibold = D \cdot \vec{F}^l_\ibold \hbox{ , }
\ibold \in \Omega^l_{valid}.
\end{gather}

It is useful to
express $D^{comp}$ as the application of the level
divergence operator $D$ applied to
extensions of $\vec{F}^{l, comp}$ to the entire level, followed by a
step that corrects the cells in $\Omega^l_{valid}$ that are
adjacent to $\Omega^{l+1}$.
We define a {\it flux register} $\delta \vec{F}^{l+1}$ associated with the fine
level
\begin{gather*}
\delta \vec{F}^{l+1} = (\delta F^{l+1}_0, ..., \delta F^{l+1}_{\Dim-1})
\\
\delta F^{l+1}_d:{\zeta}^{l+1}_{d,+} \cup {\zeta}^{l+1}_{d,-}
\rightarrow {\mathbb{R}}^m.
\end{gather*}
Let $\vec{F}^l$ be {\it any}
coarse level vector field that extends $\vec{F}^{l, comp}$, i.e.
\begin{gather*}
F^l_d = F^{l, comp}_d \mbox{ on } \Omega^{l, \ebold^d}_{valid}.
\end{gather*}
Then, for $\ibold \in \Omega^l_{valid}$,
\begin{equation} \label{eqn:flux2}
D^{comp}(\vec{F}^{l+1,comp},\vec{F}^{l,comp})_\ibold = (D \vec{F}^l)_\ibold +
D_R(\delta \vec{F}^{l+1})_\ibold,
\end{equation}
where $\delta \vec{F}^{l+1}$ is a flux register set to be
\begin{gather*}
\delta F^{l+1}_d = \langle F^{l+1}_d\rangle  - F^l_d \mbox{ on } {\zeta}^l_{d, +}
\cup {\zeta}^l_{d, -}
\end{gather*}
and $D_R$ is the reflux divergence operator, given by the following for
valid coarse level cells adjacent to $\Omega^{l+1}$:
\begin{gather*}
D_R(\delta \vec{F}^{l+1})_\ibold = \frac{1}{h^l} \sum^{\Dim-1}_{d=0}
\sum_{
\substack
{\pm = +,-: \\ \ibold \pm \half \ebold^d \in
{\zeta}^{l+1}_{d, \mp}
}} \pm \delta F^{l+1}_{d,\ibold \pm \half \ebold^d}.
\end{gather*}
For the remaining cells in $\Omega^l_{valid}$, $D_R(\delta \vec{F}^{l+1})$
is defined to be identically zero. 

We can use this notation to define the discretizations of Poisson's 
equation we will be using to compute self-gravity. 
On a single level,
$\Omega^l$, we define $\Delta^l$, the discrete Laplacian, to be the 
standard $2\Dim + 1$ point operator, with the values used on ghost 
cells computed using quadratic interpolation:
\begin{gather}
\Delta^l \phi^l = D \cdot \vec{F}^{l,\phi} \\
F^{l,\phi}_d = \frac{\phi_{\ibold + \ebold^d} - \phi_\ibold}{h^l} \\
\phi^l_\ibold = \mathcal{I}(\phi^l, \phi^{l-1})_\ibold \hbox{ for }
\ibold \in \partial\Omega^l 
\end{gather}
where $\partial\Omega^l\equiv\mathcal{G}(\Omega^l,1) - \Omega^l$.
Here the interpolation function $\mathcal{I}$ is an $O(h^3)$ estimate
of the value on the ghost cell obtained from interpolating from values
of $\phi^{l-1}$ on $\Omega^{l-1}_{valid}$ and from the values of
$\phi^l$ on $\Omega^l$; for details, see \cite{minion96b}.  We can
then define the composite Laplacian $\Delta^{comp}$ applied to all of
the valid data on all levels, in terms of that operator and refluxing
operations.
\begin{gather}
(\Delta^{comp,l} \phi)_\ibold = (\Delta^l \phi^{l,ext})_\ibold  + 
D_R(\delta \vec{F}^{l,\phi}) \hbox{ for } \ibold \in \Omega^l_{valid}
\\
\delta \vec{F}^{l,\phi} = \langle \vec{F}^{l+1,\phi}\rangle  - \vec{F}^{l,\phi}
\end{gather}
where $\phi^{l,ext}$ is some extension of $\phi^{l,comp}$ to all of $\Omega^l$.
The resulting operator depends only on the valid values of $\phi$
in the grid hierarchy (modulo roundoff considerations; cf. 
Ref.~\cite{macogr06} for further details).

\subsection{AMR for Compressible Flows with Self-Gravity} \label{amrWithSG.sec}

The starting point for this work is the algorithm described in
\cite{bergercolella89} for solving hyperbolic conservation laws on
nested refined grids.  For the case of two levels, we assume that the
solution on both levels is known at time $t^c$.  The basic steps
to evolve the solution on both levels to time $t^c+\Delta t^c$ can
be summarized as follows:

\begin{enumerate}
\item Update the solution on the coarse grid:
\begin{gather}
U^{c}(t^c + \Delta t^c) = U^c(t^c) - \Delta t^c (D \cdot \vec{F}^c ) 
+ \Delta t^c S(U^c) \hbox{ on } \Omega^c .
\label{eqn:coarseConsUpdate}
\end{gather}
Here $\mathcal{S}(U^c)$ is computed as in Eq.~(\ref{eqn:source}) (with
the body force, $\force$, set to zero), and the discrete fluxes $\vec{F}$ are
local functions of $U^c$. We also initialize flux registers associated
with $\Omega^f$ using the same fluxes
$$
\delta \vec{F}^f = - \vec{F}^c.
$$
\item Advance the solution from $t^f$ to $t^f + \Delta t^f$ on the
  fine grid $n_{ref}$ times, $n_{ref}\Delta t^f=\Delta t^c$:
\begin{gather}
U^{f}(t^f + \Delta t^f) = U^f(t^f) - \Delta t^f (D \cdot \vec{F}^f) 
+\Delta t S(U^f) \hbox{ on } \Omega^f
\label{eqn:fineConsUpdate}
\\
\delta \vec{F}^f\, +\!\!=\, \frac{1}{n_{ref}}\langle \vec{F}^f \rangle
\nonumber
\\ t^f\, +\!\!= \,\Delta t^f.
\nonumber
\end{gather}
Any values required to compute the stencil that are contained in
$\Gamma_f - \Omega^f$ are computed by interpolating the coarse grid
values $U^c(t^c)$, $U^c(t^c + \Delta t^c)$, using linear interpolation
in time, and piecewise linear interpolation in space.
\item Synchronize the values at the old and new times:
\begin{gather}
U^{c}(t + \Delta t^c) = \langle U^{f}(t + \Delta t^c) \rangle \hbox{ on } \mathcal{C}_{n_{ref}}(\Omega^f )
\\
U^{c}(t + \Delta t^c)\, +\!\!= \,D_R(\delta \vec{F} ) \nonumber
\end{gather}
where $ \langle\cdot \rangle_\ibold$ denotes the arithmetic average onto the coarse
cell $\ibold$ of all of the values defined on fine grid cells
contained in $\ibold$.
\end{enumerate}
To extend this algorithm to the case of self-gravity, we must solve
the Poisson's equation for the gravitational potential due to the mass
distribution of the fluid on the coarse and fine levels.  As usual the
coarse level is advanced first, and the solution at $t^c+\Delta t^c$
is used to provide time interpolated boundary conditions for the fine
level at intermediate time steps. In the case of hyperbolic equations
the finite characteristic speeds ensure that a fully consistent
multilevel solution can be recovered at synchronization time with the
refluxing operation. Because of its elliptic nature, however, in order
to preserve its multilevel character Poisson's equation should be
solved simultaneously on all levels.  Notice that the coupling among
levels is enforced by the continuity of the potential (Dirichlet) and
of its normal derivative (Neumann) at the coarse/fine grid interface.
Therefore, to maintain the multilevel character of Poisson's equation
when the levels are not synchronized yet, we obtain a single level
solution to Poisson's equation on the coarse level and apply a {\it
  lagged} estimate of the effect of the coarse/fine matching
conditions at refinement boundaries, following the ideas developed
in~\cite{almgrenETAL:1998,martinColella:2000,fisherPhD2002} for
incompressible fluids.  This leads to the following modifications to
the algorithm given above.

\begin{description}
\item{0.} At simulation start, when all levels are synchronized,
we compute a composite grid solution to Poisson's equation
\begin{eqnarray}
(L^{comp,c}\phi^{comp})(t^c)_\ibold & = & \rho^c(t^c)_\ibold \quad 
\ibold\in\Omega^c_{valid} \label{eqn:phiCompCrse}
\\ \label{eqn:phiCompFine}
(L^{comp,f}\phi^{comp})(t^c)_\ibold & = & \rho^f(t^c)_\ibold \quad \ibold\in\Omega^f
\end{eqnarray}
as well as the coarse grid solution
\begin{equation}
(L^c \phi^c)(t^c) = \rho^c(t^c) \hbox{ on } \Omega_c .
\label{eqn:crsGrid}
\end{equation}

Here, and in what follows, we will denote by
$L(t)\equiv\frac{a(t)}{4\pi G}\Delta$ with superscripts $l$, $c$, $f$,
$comp$ indicating the particular discretization of the Laplacian
operator.
\item{1.} Together with $U(t^c)$, 
the acceleration $\force^c(t)$ derived from the
  composite solution of the potential on the coarse grid is used to
  compute the coarse grid fluxes and source terms. After updating the conserved
  quantities using (\ref{eqn:coarseConsUpdate}), we compute the
  coarse grid potential at the new time
\begin{gather} \label{eqn:singleLevPot}
(L^c \phi^c) (t^c+\Delta t^c)  = \rho^c(t^c + \Delta t^c) \\
\tilde{\phi}^{c,comp} (t^c+\Delta t^c)  = \phi^c (t^c+\Delta t^c)+
(\phi^{c,comp}(t^c)-\phi^c(t^c))  \label{eqn:approxCompPot}
\end{gather}
where in the latter step we have approximately corrected the coarse grid 
single-level-solution of the potential for the effects due to the
solution on the finer grid~\cite{almgrenETAL:1998,martinColella:2000}.  We
use the solution in Eq.~(\ref{eqn:approxCompPot}) with
$\phi^{c,comp}(t^c)$ to obtain boundary conditions interpolated in
time for the potential at the fine level at intermediate timesteps.
\item{2.} We apply the update (\ref{eqn:coarseConsUpdate}) on the 
finer level. At $t^f=t^c$ in order to
compute the hydrodynamic fluxes and source terms we use
the force, $\force^f$, derived from 
the composite potential solution to Eq.~(\ref{eqn:phiCompFine}).
At intermediate steps, $t^c<t^f < t^c +\Delta t^c$, 
we solve the following Poisson's equation on the fine grid 
with interpolated boundary conditions
\begin{gather}
(L^f \phi^f)(t^f) = \rho^f (t^f) \hbox{ on } \Omega^f
\label{eqn:finePoisson}
\\ \nonumber
\phi^f(t^f) = \mathcal{I}[\phi^f(t^f),\tilde{\phi}^{c}(t^f)] \hbox{ on } \partial \Omega^f 
\end{gather}
where $\tilde{\phi}^{c}(t^f)$
is obtained by linear interpolation in time as
\begin{gather*}
\tilde{\phi}^c(t^f) = (1-\alpha)\, \phi^{c,comp}(t^c) 
+ \alpha\, \tilde{\phi}^{c,comp} (t^c+\Delta t^c) 
\text{ , } \alpha=\frac{t^f-t^c}{\Delta t^c}.
\end{gather*}
We compute the forces at the new timestep and apply the correction
(\ref{eqn:forceUpdate})-(\ref{eqn:ds}) to the fluid momentum and kinetic energy.
\item{3.} At time of synchronization, $t^f=t^c+\Delta t^c$, we solve
  the set of equations (\ref{eqn:phiCompCrse})-(\ref{eqn:crsGrid}),
  for the composite and single level solution of the potential.  We
  use the composite solution to derive the new force, $\force^{n+1}$,
  and apply momentum and kinetic energy corrections
  (\ref{eqn:forceUpdate})-(\ref{eqn:ds}) to obtain time centered
  forces on both coarse and fine levels. The flow of the calculation
  restarts from step 1 with the gravitational force known at all levels.
\end{description}

\subsection{AMR with Particles} \label{Particles.sec}
Due to the time refinement character of the AMR technique the solution
on different levels is advanced with different timesteps. This implies
that the density field represented by the particles evolved on the finer level
may not be available on the coarser level unless they are synchronized.
However, this is information is necessary to solve Poisson's equation. 
Therefore, we find it convenient to introduce {\it effective} particles
to recover such information in a computationally inexpensive way and
without compromising the code accuracy. 

Thus we introduce an {\it aggregation} operation $\mathcal{P}
\rightarrow \langle \mathcal{P} \rangle ^l$ that projects a collection
of particles covered by $\Omega^l$ onto a set of effective particles,
with no more that one particle per cell. If $p \in \langle
\mathcal{P}\rangle ^l$, then
\begin{eqnarray} \label{aggr1:eq}
m_{p}  & = & \sum \limits_{p': \xbold_{p'} \in V_\ibold} m_{p'} \\
\xbold_{p}  & = &  
\frac{1}{m_p} \sum \limits_{p': \xbold_{p'} \in V_\ibold} m_{p'} \xbold_{p'} \\
 \label{aggr3:eq}
\ubold_{p}  & = &  
\frac{1}{m_p} \sum \limits_{p': \xbold_{p'} \in V_\ibold} m_{p'} \ubold_{p'}
\end{eqnarray}
The aggregation operation conserves the
monopole and dipole terms but causes information to be lost on the
quadrupole moment of the aggregated particles~\cite{mcaa93}, which
provides corrections to the potential of order $h^2$.
Thus the aggregation step preserves second order accuracy.
Note, also, that the potential and force fields obtained through the 
aggregated particles are only used to provide boundary conditions for 
the finer level.

Restricting again the discussion to the case of two levels of refinement,
the changes in the algorithm described above are given as follows:
\begin{description}
\item{0.} At simulation start, we partition the
  particles into ones that will be evolved using the coarse and fine time
  steps. If $\mathcal{P}$ is the set of all particles,
\begin{gather} \label{eqn:partition0}
\mathcal{P}^f = \big\{p \in \mathcal{P} : \xbold_p \in \mathcal{G}(\Omega^f, -n_{ref} n_{buf})\big\}
\\ \label{eqn:partition1}
\mathcal{P}^c = \mathcal{P} - \mathcal{P}^f .
\end{gather} 

The parameter $n_{buf}$ is chosen so that the support for the
Particle-Mesh interpolation function, used to calculate the force
acting on the particle and the particle mass distribution on the grid,
is completely contained in $\Omega^f$ for all of the fine grid time
steps, $n_{ref}\Delta t^f=\Delta t^c$. Using $n_{buf}=1$ and $C_{part}=0.5$
is sufficient for all choices of Particle-Mesh scheme used here (see Appendix~\ref{cas.ap}).

We then define the set $\langle\mathcal{P}^{f}\rangle^{c}$ of fine
particles aggregated on the coarse grid using
Eq.~(\ref{aggr1:eq})-(\ref{aggr3:eq}).

Finally, in computing the gravitational potential the densities in
Eq.~(\ref{eqn:phiCompCrse})-(\ref{eqn:crsGrid}) are modified to
account for the mass distribution of the particles:
\begin{gather}
\rho_\ibold ^c(t^c) =  \rho_\ibold ^{c,fluid}(t^c)    +  
\sum \limits_{p \in \mathcal{P}^c \cup \langle \mathcal{P}^f\rangle ^c}
m_p W\Bigg( \frac{(\ibold+\frac{1}{2}\ubold)\,h^c-\xbold_p(t^c)}{h^c}\Bigg)
\label{eqn:dmp1}
\\
\begin{split}
\rho_\ibold^f(t^c) =  \rho_\ibold^{f,fluid}(t^c)  & +  
\sum \limits_{p \in \mathcal{P}^f } m_p W\Bigg(\frac{(\ibold+\frac{1}{2}\ubold)\,h^f-\xbold_p(t^c)}{h^f}\Bigg) 
\\
 &  +   \sum \limits_{p \in \mathcal{P}^c}
m_p W\Bigg( \frac{(\ibold+\frac{1}{2}\ubold)\,h^f-x_p(t^c)}{h^c}\Bigg).
\end{split}
\label{eqn:dmp2}
\end{gather}
Here $W$ is one of the Particle-Mesh assignment schemes used to spread
the particle mass on the grid, described in Appendix~\ref{cas.ap}.
Note that the addition to $\rho^c$ of the density field due to
$\langle \mathcal{P}^f\rangle ^c$ has no effect on the composite
solution, because the support of $W$ for each particle $\in \langle
\mathcal{P}^f\rangle ^c$ is contained entirely in
$\mathcal{C}_{n_{ref}}(\Omega^f)$.
\item{1.}  The gravitational force resulting from the composite
  solution of the potential is used to compute both the fluid fluxes
  and source terms, as well as to perform the update
  (\ref{eqn:particleVelocityPredictor})-(\ref{eqn:particlePositionPredictor})
  for the particles in $\mathcal{P}^c \cup \langle
  \mathcal{P}^f\rangle ^c$. For the latter, we interpolate the
  accelerations from the grid cells to particle positions with one
  of the methods described in Appendix~\ref{cas.ap}.
\item{2.}  At the end of each fine time step, while
 $t^f + \Delta t^f< t^c+\Delta t^c$, we compute the new fine potential,
  $\phi^f$, modifying the mass density as
\begin{gather}
\begin{split}
\rho^f(t^f)_\ibold = \rho^{f,fluid}(t^f)_\ibold & + 
\sum \limits_{p \in \mathcal{P}^f }
m_p W\Bigg(\frac{(\ibold+\frac{1}{2}\ubold)\, h^f-\xbold_p(t^f)}{h^f}\Bigg)
\\
& + \sum \limits_{p \in \mathcal{P}^c}
m_p W\Bigg(\frac{(\ibold+\frac{1}{2}\ubold)\, h^f-\xbold_p(t^f)}{h^c}\Bigg) \hbox{ , }\ibold \in \Omega^f
\end{split}
\label{eqn:finerhs}
\end{gather}
where the positions of the particles in $\mathcal{P}^c$ at the
intermediate times are given by linear interpolation between
$\xbold_p(t^c)$ and $\xbold_p(t^c+\Delta t^c)$. 
  We then use the acceleration due
to this field to update the fine particle velocities using
(\ref{kdku.eq}), and the fine fluid state using
(\ref{eqn:forceUpdate})-(\ref{eqn:ds}).
\item{3.} The synchronization step is analogous to the one in
  Sec.~\ref{amrWithSG.sec}: we calculate a single grid and composite
  grid solution of the potential using the total mass density
  distribution of fluid and particles given by
  Eq.~(\ref{eqn:dmp1})-(\ref{eqn:dmp2}).  The gravitational force
  derived from the composite potential is used to apply the corrections to
  the particle velocity, given in Eq.~(\ref{kdku.eq}), and to the fluid
  momentum and kinetic energy, given in Eq.~(\ref{eqn:forceUpdate})-(\ref{eqn:ds}), at
  all levels. Finally, the sets $\mathcal{P}^c,~\mathcal{P}^f$ and
  $\langle \mathcal{P}^f\rangle ^c$ are upgraded following the definitions
in (\ref{eqn:partition0})-(\ref{eqn:partition1}) to account for the
  new particle positions. The flow of the calculation restarts from
  step 1.
\end{description}

\subsection{The General Multilevel Algorithm}

For the case of a general $(l_{max}+1)$-level calculation,
we assume that at simulation start all the particles have
been partitioned into groups corresponding to the levels
on which they shall be advanced, and that the particles being
advanced by finer grids have been aggregated into effective particles
for the next coarser levels. This is summarized in the procedure:
\begin{algorithmic}
\FOR {$l=0, \dots l_{max}-1$}
\STATE
{$\mathcal{P}^{l+1}=\big\{\xbold\in \mathcal{P}^{l}:\xbold_p(t^l)\in \mathcal{G}(\Omega^{l+1}, -n_{ref}n_{buf})\big\}$}
\STATE
{$\mathcal{P}^{l}\leftarrow \mathcal{P}^{l}-\mathcal{P}^{l+1}$}
\ENDFOR
\STATE $\mathcal{P}^{l_{max}}_c=\emptyset$
\FOR
{$l=l_{max}-1,\dots , 0$} \STATE
{$\mathcal{P}^{l}_c = \langle \mathcal{P}^{l+1}\cup\mathcal{P}^{l+1}_c\rangle ^{l}$}
\ENDFOR
\end{algorithmic}

We can then describe 
the algorithm for {\bf advance}$(l)$ that advances the solution at
level $l$, $0\leq l\leq l_{max}$ by one time step.  We assume that the
solution is known at time $t^l+\Delta t^l$ in a process that includes
a recursive application of {\bf advance}.  At the beginning of {\bf
  advance}, we assume that: we know the fluid state $U^l$ and particle
state $P^l$ associated with that level at time $t^l$; we know the
fluid, particle, and potential at the next coarse level at times
$t^{l-1}$, $t^{l-1}+\Delta t^{l-1}$; a composite solution as well as
single level solution of the gravitational potential at time $t^l$ has
been computed on level $l-1$ and on all finer levels;

We then advance the solution by a time step $\Delta t^l$, with
$n^{l-1}_{ref}\Delta t^l=\Delta t^{l-1}$, as follows:

\begin{enumerate}

\item Using $U(t^l)$ along with the accelerations, $\force^l_\ibold$, 
$\ibold\in\Omega^l$, computed from $\phi^l$, we calculate the hyperbolic 
fluxes $\vec{F}^l$ and the source terms $S(t^l)$. 
We update the conserved quantities
\begin{gather}
U(t^l+\Delta t^l)=U(t^l)-\Delta t^lD\cdot\vec{F}^l+\Delta t^l S^l(t^l)
\\
\delta F^{l+1}=-\vec{F}^l \text{ if } l<l_{max} \nonumber
\\
\delta F^{l}\, +\!\!= \, \langle \vec{F}^l\rangle  \text{ if } l > 0 \nonumber
\end{gather}
and advance the positions and velocities of the particles in
$\mathcal{P}^l \cup \mathcal{P}^l_c$ using
(\ref{eqn:particleVelocityPredictor}) and
(\ref{eqn:particlePositionPredictor}).
\item With $t^l\leftarrow t^l+\Delta t^l$,
we solve Poisson's equation on level $l$
\begin{align}
& (L^l\phi^l)(t^l)_\ibold= \rho^{fluid,\thinspace l}(t^l)_\ibold
 +\sum_{p\in p^l} m_p W\Bigg(\frac{(\ibold+\frac{1}{2}\ubold)\,h^l-\xbold_p(t^l)}{h^l}\Bigg) \label{eqn:levelLPoisson}
 ~~~~~~~~~~~~~~~~~ \\ +& \sum_{p\in p^l_c} m_p W\Bigg(\frac{(\ibold+\frac{1}{2}\ubold)\,h^l-\xbold_p(t^l)}{h^c}\Bigg) 
 \nonumber +\sum_{p\in p^{l-1}} m_p W\Bigg(\frac{(\ibold+\frac{1}{2}\ubold)\,h^l-x_p(t^l)}{h^{l-1}}\Bigg)
\hbox{ , } \ibold\in\Omega^l
\end{align}
\begin{gather}
\phi^l(t^l)_\ibold=\mathcal{I}[\phi^l(t^l),
\tilde{\phi}^{l-1}(t^{l}) ]_\ibold,\thinspace \ibold\in\partial\Omega^l \nonumber
\end{gather}
where $\tilde\phi^{l-1}(t)$ is obtained by linear interpolation in
time between $\tilde\phi^{l-1,comp}(t^{l-1})$ and
$\tilde{\phi}^{l-1,comp}(t^{l-1}+\Delta t^{l-1})$ with the latter
defined in
Eq.~(\ref{eqn:approxCompPot}).  As in Eq.~(\ref{eqn:approxCompPot}) we
then estimate
\begin{equation}
\tilde{\phi}^{l,comp} (t^l) = \phi^l (t^l)+
(\phi^{l,comp}(t^{l-1})-\phi^l(t^{l-1}))
\end{equation}
which is used to provide boundary conditions for the potential at the
next finer level, $\phi^{l+1}(t)$.
\item
We call advance recursively $n^l_{ref}$ times.
\begin{algorithmic}
\STATE $t^{l+1}=t^l$
\WHILE {$t^{l+1} < t^l + \Delta t^l$} \STATE
{$\text{\bf advance}(l+1)$}
\STATE {$t^{l+1} = t^{l+1} + \Delta t^{l+1}$}
\ENDWHILE
\end{algorithmic}
\item  If $t^{l-1} <t^l < t^{l-1}+\Delta t^{l-1}$, 
 we obtain $\{\phi^{l',comp}(t^l)\}_{l'\geq l}$ by 
 solving the composite equations :
\begin{align}
(L^{comp,l'}\phi^{comp})(t^l)_\ibold =  \rho^{fluid,l'}(t^l)_\ibold
& +\sum_{p\in \mathcal{P}^{l'}} m_p W\Bigg(\frac{(\ibold+\half\ubold)h^{l'}-x_p(t^l)}{h^{l'}}\Bigg)
\\
& +\sum_{p\in \mathcal{P}^{l'-1}} m_p W\Bigg(\frac{(\ibold+\half\ubold)h^{l'}-x_p(t^l)}{h^{l'-1}}\Bigg) \nonumber
\end{align}
\begin{gather*}
\ibold\in\Omega^{l'}_{valid}\quad ,\quad l'=l,...,l_{max}
\end{gather*}
\begin{gather*}
\phi^{l,comp}(t^l)_\ibold=\mathcal{I}[\phi^{l,comp}(t^l),
\tilde\phi^{l-1}(t^{l}) ]_\ibold,\thinspace \ibold\in\partial\Omega^l .
\end{gather*}
The field so obtained is used to compute accelerations at the new time,
and update the fluid and particle velocities using
(\ref{eqn:forceUpdate}) - (\ref{eqn:ds}) and (\ref{kdku.eq}) respectively.

\item The solution at level $l$ is synchronized with the solutions at the
  finer levels. 
\begin{gather*}
U(t^l + \Delta t^l) = \langle U(t^{l+1})\rangle  \text{ on } \mathcal{C}_{n^l_{ref}}(\Omega^{l+1})
\\
U(t^l + \Delta t^l)\, +\!\!= \, D_R(\delta \vec{F}^{l+1})
\end{gather*} 
We upgrade the sets $\mathcal{P}^{l'}$ for $ l'=l-1,...,l_{max}$ and
$\langle \mathcal{P}^{l'+1}\rangle ^{l'}$ for $l'=l-1,...,l_{max}-1$,
according to the new particle positions. The flow of the calculation
restarts from step 1.
\end{enumerate}
\section{Convergence Tests} \label{tests.sec}
We have implemented the above schemes in a Cosmological Hydromagnetic
AMR Radiation Many-body (CHARM) code. The code is based on the CHOMBO
AMR library and it is implemented in a hybrid C++/Fortran77 language.
Additional physics modules, such as radiation~\cite{mico06b},
cosmic-rays~\cite{min01,min06} and magnetohydrodynamics will be presented
elsewhere. In the following, we focus on numerical tests to assess the
performance of the algorithms in terms of accuracy and applicability
to problems of direct interest. Performance tests will be presented
elsewhere.

Unless explicitly stated otherwise, in the following we use these CFL
coefficients for the time step: $C_{\rm hydro}=C_{\rm part} =0.5$ and
$C_{\rm exp}=0.01$. In addition, we restrict the results to the case of a 
TSC interpolation scheme which, in accord with previous authors, we 
find to give the most accurate results. 
\begin{table} [ht]
\label{t1:tab}
\caption{Convergence tests: collisionless case$^\dagger$ }
\begin{tabular*}{\textwidth}{@{\extracolsep{\fill}}lcccccc}
\hline 
\hline 
$N_{\rm part}$ & $L_1$ & $R_1$ & $L_2$ & $R_2$ & $L_\infty$ & $R_\infty$ \cr
\hline 
\multicolumn{7}{c}{\bf position}  \cr
8    &  1.3e-07 & 1.9 & 1.4e-07 & 1.9  & 1.9e-07 & 1.8   \cr
16   &  3.5e-08 & 2.0 & 3.9e-08 & 2.0  & 5.4e-08 & 2.0   \cr
32   &  8.9e-09 & 2.0 & 9.8e-09 & 2.0  & 1.3e-08 & 2.1   \cr
64   &  2.2e-09 & 2.0 & 2.4e-09 & 1.9  & 3.0e-09 & 1.6   \cr
128  &  5.6e-10 &  -- & 6.2e-10 &  --  & 9.8e-10 &  --   \cr
\hline
\multicolumn{7}{c}{\bf velocity}  \cr
8    &  1.3e-04 & 1.9 & 1.4e-04 & 1.9  & 1.9e-04 & 1.8   \cr
16   &  3.5e-05 & 2.0 & 3.8e-05 & 2.0  & 5.3e-05 & 2.0   \cr
32   &  8.8e-06 & 2.0 & 9.7e-06 & 2.0  & 1.3e-05 & 2.1   \cr
64   &  2.2e-06 & 2.0 & 2.4e-06 & 1.9  & 3.0e-06 & 1.6   \cr
128  &  5.5e-07 & --  & 6.2e-07 & --   & 9.7e-07 & --    \cr
\hline
\multicolumn{7}{c}{\bf force}  \cr
8    &  6.5e-02 & 1.9 & 7.1e-02  & 1.9  & 9.4e-02  & 1.8 \cr
16   &  1.7e-02 & 2.0 & 1.9e-02  & 2.0  & 2.7e-02  & 2.0 \cr
32   &  4.4e-03 & 2.0 & 4.8e-03  & 2.0  & 6.5e-03  & 2.1 \cr
64   &  1.1e-03 & 2.0 & 1.2e-03  & 1.9  & 1.5e-03  & 1.6 \cr
128  &  2.8e-04 & --  & 3.1e-04  & --   & 4.8e-04  & --  \cr
\hline
\hline 
\end{tabular*}
 \tablenotes{
 \qquad\llap{$^\dagger$} We use equal number of cells and particles, $N_{\rm part}=N_{\rm cell}$, a cell-centered force scheme, and a constant $\frac{\Delta t}{\Delta x} = 1.6\times 10^{-4}$.
 }
\end{table}
Errors and convergence rates are calculated as follows.  At a given
resolution, $r$, for any given cell or particle, ${\bf i}$, we
estimate the error on a {\it computed} quantity, $q^c_r({\bf i})$, with
respect to the {\it analytic} solution,$q^a({\bf i})$, as
\begin{equation} \label{numerr:eq}
\delta q_r({\bf i}) = q^c_r({\bf i}) - q^a_r({\bf i}).
\end{equation}
We then compute the n-norm of the error, i.e.
\begin{equation} \label{lnorm_n:eq}
L_n(\delta q_r) = \| \delta q_{r} \|_n =  \left[\sum |\delta q_r({\bf i})|^n  v_{\bf i}\right]^{1/n}
\end{equation}
where $v_{\bf i}$ is either the ${\bf i}$-th cell volume or the
inverse of the number of particles; finally we estimate the
convergence rate as
\begin{equation} \label{conv_rate:eq}
R_n = \frac{ \ln[L_n(\delta q_r)/L_n(\delta q_s)] }{ \ln (\Delta x_r / \Delta x_s) }.
\end{equation}
For the cases studied below we report the $L_1,~L_2$ and $L_\infty$
norm of the errors.
\subsection{Zel'dovich's Pancake} \label{zeld:sec}
We begin with a classical test problem for cosmological codes, the
evolution of a one-dimensional plane-wave perturbation in an expanding
background.  
%
\begin{figure}[ht]
\begin{center}
\includegraphics[height=0.6\textheight, scale=1.0]{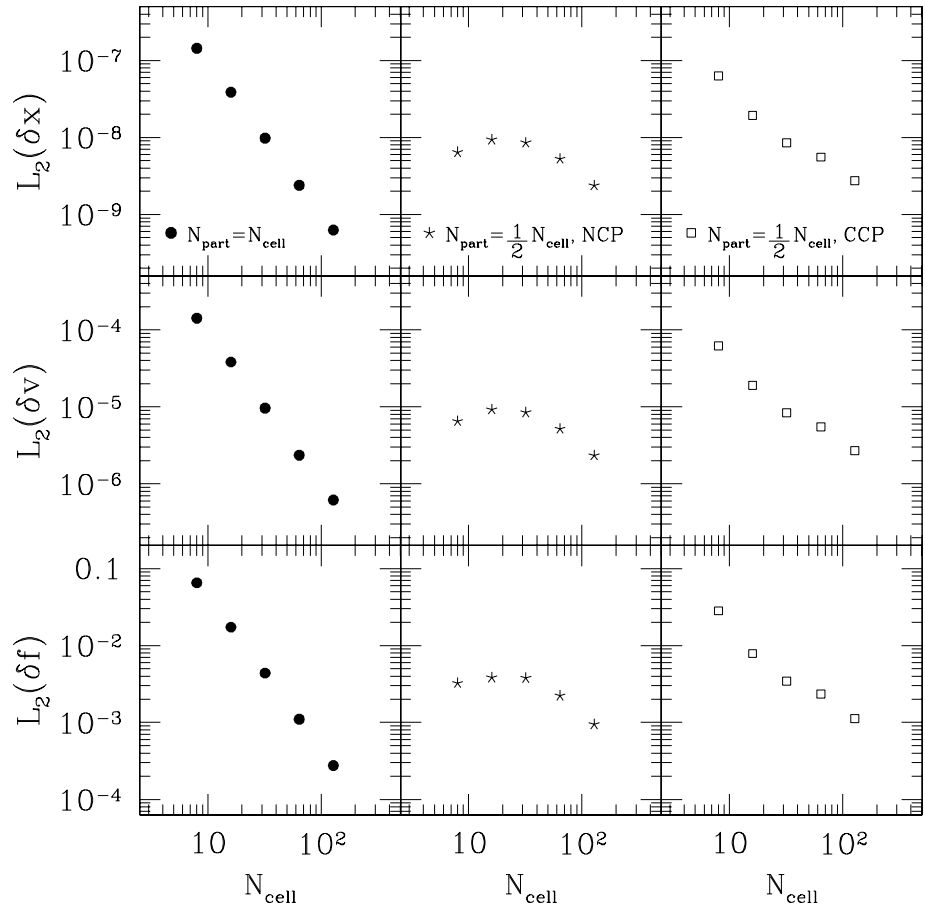}
\caption{
$L_2$ norm of the error in position (top), velocity (center) and 
force (bottom) as a function of the number of grid cells.
Left panels correspond to the case in which the number of particles 
and grid cells is the same, $N_{cell}=N_{part}$. Central and
right panels correspond to $N_{part}=\frac{1}{2}N_{cell}$, with 
particles initially placed either at cell nodes (middle panels) 
or at cell centers (right panels). In all cases a two point cell 
centered force stencil is used. See legend for the meaning of the symbols
(UG= Uniform Grid, $a$ is the expansion parameter).
\label{f1:fig}}
\end{center}
\end{figure}
In Zel'dovich's formulation~\cite{zeldovich70a}, the
comoving position and peculiar velocities of collisionless matter
evolve as
\begin{eqnarray}
x(t) & = & q + \frac{b(t)}{a(t)}\;  p(q) \\
v(t) & = & a(t)\; \dot{x} (t) 
\end{eqnarray}
where $q,p$ are the Lagrangian initial position and displacements,
$a(t)$ is the expansion factor and $b(t)/a(t)$ describes the growth
factor of the perturbation. For a closed universe ($\Omega_m=1$),
$a(t)=(3H_0t/2)^{2/3}$ (Eq.~[\ref{tepx.eq}]) and $b(t)= 2a^2
/5$~\cite{zeldovich70a}. 

\begin{table}  [ht]
\label{t2:tab}
\caption{Convergence tests: collisionless, variable timestep, linear phase$^\dagger$}
\begin{tabular*}{\textwidth}{@{\extracolsep{\fill}}lcccccc}
\hline 
\hline 
$N_{\rm part}$ & $L_1$ & $R_1$ & $L_2$ & $R_2$ & $L_\infty$ & $R_\infty$ \cr
\hline 
\multicolumn{7}{c}{\bf position}  \cr
8    &  6.4e-06 & 1.9 & 6.9e-06 & 1.9  & 9.2e-06 & 1.8   \cr
16   &  1.7e-06 & 2.0 & 1.9e-06 & 2.0  & 2.6e-06 & 2.0   \cr
32   &  4.3e-07 & 2.0 & 4.7e-07 & 2.0  & 6.4e-07 & 2.1   \cr
64   &  1.1e-07 & 2.0 & 1.1e-07 & 1.9  & 1.4e-07 & 1.6   \cr
128  &  2.5e-08 &  -- & 2.8e-08 &  --  & 4.5e-08 &  --   \cr
\hline
\multicolumn{7}{c}{\bf velocity}  \cr
8    &  8.9e-04 & 1.9 & 9.6e-04 & 1.9  & 1.3e-03 & 1.8   \cr
16   &  2.3e-04 & 2.0 & 2.6e-04 & 2.0  & 3.6e-04 & 2.0   \cr
32   &  5.9e-05 & 2.0 & 6.5e-05 & 2.0  & 8.9e-05 & 2.1   \cr
64   &  1.5e-06 & 2.0 & 1.6e-05 & 1.9  & 2.0e-05 & 1.6   \cr
128  &  3.7e-07 & --  & 4.2e-06 & --   & 6.6e-06 & --    \cr
\hline
\multicolumn{7}{c}{\bf force}  \cr
8    &  6.5e-02 & 1.9 & 7.1e-02  & 1.9  & 9.4e-02  & 1.8 \cr
16   &  1.7e-02 & 2.0 & 1.9e-02  & 2.0  & 2.7e-02  & 2.0 \cr
32   &  4.4e-03 & 2.0 & 4.8e-03  & 2.0  & 6.5e-03  & 2.1 \cr
64   &  1.1e-03 & 2.0 & 1.2e-03  & 1.9  & 1.5e-03  & 1.6 \cr
128  &  2.8e-04 & --  & 3.1e-04  & --   & 5.1e-04  & --  \cr
\hline
\hline 
\end{tabular*}
 \tablenotes{
 \qquad\llap{$^\dagger$} We use equal number of cells and particles, 
$N_{\rm part}=N_{\rm cell}$, a cell-centered force scheme, a variable 
timestep, $\frac{\Delta t}{\Delta x} = C_{exp}(\frac{a}{\dot{a}}),
~C_{exp}=10^{-2}$, and $a=0.0221$.
 }
\end{table}
Setting the initial displacements to a
sinusoidal form, $p(q)=5A\sin(kq)/2$, where $k$ is the perturbation
wavenumber, we obtain
\begin{eqnarray}
x(t) & = & q + a\;A \sin(kq) \label{pos.eq} \\
v(t) & = & a\; \dot{a}\;A \sin(kq) \label{vel.eq} \\
\rho(t) & = & \rho_0 \;  \left[ 1+a\; A\; k\;\cos(kq) \right]^{-1}.
\label{rho.eq}
\end{eqnarray}
The solution described by Eq.~(\ref{pos.eq}) becomes 
singular, that is $\partial x/\partial q = 0$ when $a_{\rm collapse}=(Ak)^{-1}$.
At this stage, particles trajectories cross at $x=q=\pi/k$ and a caustic forms.
In the following we use $a_{\rm start}=1/51$, 
$a_{\rm collapse}=1/2$ and $k=2\pi/h^{-1}L_{\rm box}$.
\subsubsection{Collisionless Component}
Table~\ref{t1:tab} demonstrates the second order accuracy of the code.  
The different columns report, as a function of numerical resolution, the
$L_1,~L_2$ and $L_\infty$ norms of the error on the particles
position, velocity and force, and the corresponding convergence rates
of the errors. The $L_2$ norm of these errors is also reported
graphically in the left hand side panels of Fig.~\ref{f1:fig}. For this case we
use equal number of cells and particles, $N_{\rm part}=N_{\rm
  cell}$, a cell-centered force scheme, and a constant $\frac{\Delta
  t}{\Delta x} = 1.6\times 10^{-4}$.
\begin{table} [ht]
\label{t3:tab}
\caption{Convergence tests: collisionless, variable timestep, nonlinear phase$^\dagger$}
\begin{tabular*}{\textwidth}{@{\extracolsep{\fill}}lccccccccc}
\hline
\hline
 & \multicolumn{4}{c}{uniform grid}  & \multicolumn{5}{c}{AMR} \cr
 \cline{2-5} \cline{6-10}
$N_{\rm part}$ & $L_1$ & $R_1$ & $L_\infty$ & $R_\infty$ 
 & $L_1$ & $R_1$ & $L_\infty$ & $R_\infty$ & $l_{max}$ \cr
\hline
 & \multicolumn{8}{c}{\bf position}  \cr
8    &  3.0e-02 & 1.7 & 4.6e-02 & 1.6   &  2.9e-02 & 1.9 & 4.5e-02 & 1.9 & 1\cr
16   &  9.5e-03 & 1.8 & 1.6e-02 & 1.8   &  7.9e-03 & 2.0 & 1.2e-02 & 2.0 & 1\cr
32   &  2.7e-03 & 1.9 & 4.7e-03 & 1.8   &  2.0e-03 & 2.0 & 3.1e-03 & 1.9 & 2\cr
64   &  7.2e-04 & 2.0 & 1.3e-03 & 2.0   &  5.1e-04 & 2.0 & 8.3e-03 & 1.9 & 2\cr
128  &  1.8e-04 & 2.0 & 3.3e-04 & 1.8   &  1.2e-04 & 2.0 & 2.2e-04 & 1.3 & 3\cr
256  &  4.6e-05 &  -- & 9.2e-05 &  --   &  3.0e-05 &  -- & 9.2e-05 &     & 3\cr
\hline
 & \multicolumn{8}{c}{\bf velocity}  \cr
 8    &  5.9e-02 & 1.4 & 9.4e-02 & 1.2 &  5.1e-02 & 1.8 & 7.9e-02 & 1.9 & 1\cr
 16   &  2.2e-02 & 1.6 & 4.1e-02 & 1.2 &  1.4e-02 & 1.8 & 2.1e-02 & 1.5 & 1\cr
 32   &  7.4e-03 & 1.6 & 1.8e-02 & 1.3 &  4.1e-03 & 2.1 & 7.4e-03 & 2.2 & 2\cr
 64   &  2.4e-03 & 1.7 & 7.3e-03 & 1.4 &  9.6e-03 & 2.1 & 1.6e-03 & 1.9 & 2\cr
 128  &  7.4e-04 & 1.8 & 2.8e-03 & 1.6 &  2.2e-04 & 2.0 & 4.5e-04 & 1.3 & 3\cr
 256  &  2.1e-04 & --  & 9.1e-04 & --  &  5.7e-05 & --  & 1.8e-04 & --  & 3\cr
 \hline
 & \multicolumn{8}{c}{\bf force}  \cr
 8    &  1.5e-01 & 1.1 & 2.6e-01  & 0.7 & 1.1e-01 & 1.3 & 1.8e-01 & 0.9 & 1\cr
 16   &  7.4e-02 & 1.0 & 1.6e-01  & 0.5 & 4.5e-02 & 2.4 & 9.3e-02 & 2.6 & 1\cr
 32   &  3.7e-02 & 1.0 & 1.2e-01  & 0.6 & 8.2e-02 & 1.2 & 1.6e-02 & 0.1 & 2\cr
 64   &  1.8e-02 & 1.2 & 7.8e-02  & 0.7 & 3.5e-03 & 0.5 & 1.5e-02 & 2.3 & 2\cr
 128  &  8.1e-03 & 1.4 & 4.6e-02  & 1.0 & 2.5e-03 & 1.6 & 3.0e-02 & 1.3 & 3\cr
 256  &  3.1e-03 & --  & 2.2e-02  & --  & 8.2e-04 & -- & 1.2e-02  & --  & 3\cr
 \hline
\hline 
\end{tabular*}
 \tablenotes{
 \qquad\llap{$^\dagger$} We use equal number of cells and particles, 
$N_{\rm part}=N_{\rm cell}$, a cell-centered force scheme, a variable 
timestep, $\frac{\Delta t}{\Delta x} = C_{exp}(\frac{a}{\dot{a}}),
~C_{exp}=10^{-2}$, and $a=0.479$.
 }
\end{table}
In Fig.~\ref{f1:fig} we also report the $L_2$ norm of the errors
for the case in which $N_{\rm part}=\frac{1}{2}N_{\rm cell}$ and
the particles are initially placed either at cell nodes 
(central panels, NCP for node centered particle) or cell
centers (right panels, CCP for cell centered particle).
Both these configurations lead to a slower convergence rate than
our reference case illustrated in the left hand side panels.

\begin{figure} [ht]
\begin{center}
\includegraphics[height=0.6\textheight, scale=1.0]{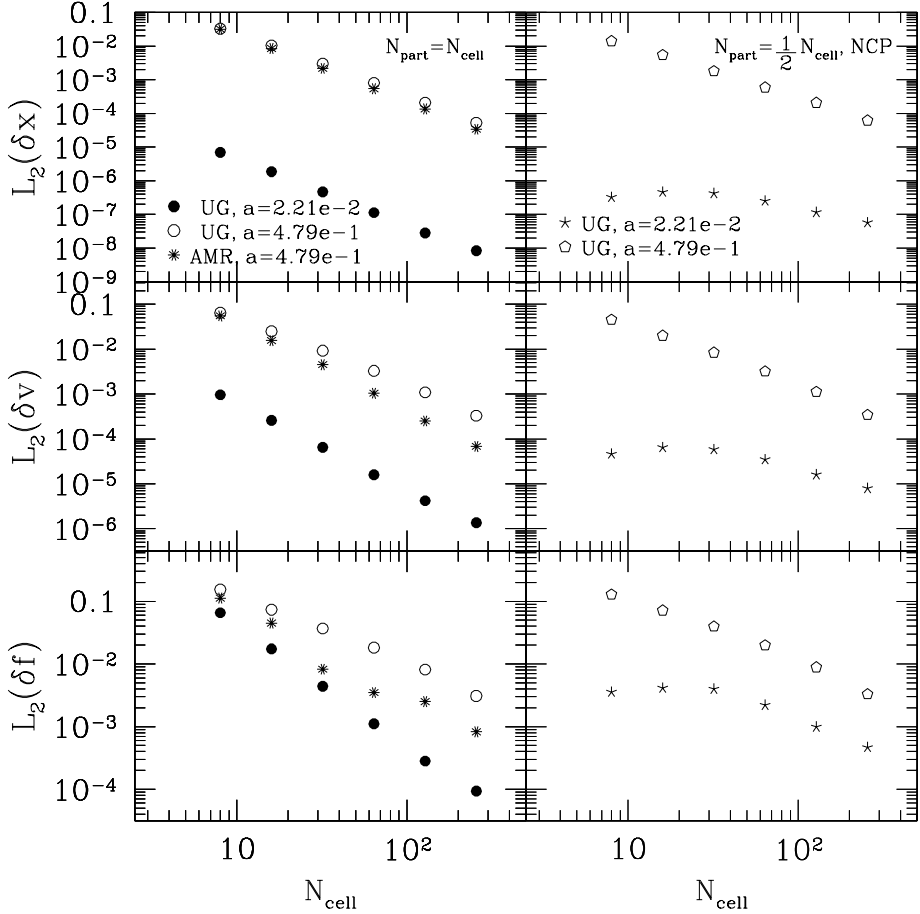}
\caption{
$L_2$ norm of the error in position (top), velocity (center) and 
force (bottom) as a function of the number of grid cells. Left 
panels correspond to the case in which the number of particles and grid
cells is the same, $N_{cell}=N_{part}$. Right panels correspond to 
$N_{part}=\frac{1}{2}N_{cell}$, with particles initially placed 
at cell nodes. See legend for the meaning of the symbols
(UG= Uniform Grid, $a$ is the expansion parameter).
\label{f2:fig}}
\end{center}
\end{figure}

In cosmological simulations, the timestep during the initial stages is
determined by the expansion rate of the background. Thus in
Table~\ref{t2:tab} we report the same quantities as in
Table~\ref{t1:tab} but for the case of a (variable) timestep
determined by $\frac{\Delta t}{\Delta x} =
C_{exp}(\frac{a}{\dot{a}})$, with $C_{exp}=10^{-2}$.  The $L_2$ norm
of these errors are also reported as filled symbols in the left hand
side panels of Fig.~\ref{f2:fig}.  The errors were computed after ten
timesteps so that the system is still in the linear regime. And, in
fact, the convergence rates are the same as in Table~\ref{t1:tab}.
The same correspondence in terms of convergence rates also exists for
the case in which $N_{\rm part}=\frac{1}{2}N_{\rm cell}$, as illustrated
for the NCP case by the star symbols in right hand side panels of
Fig.~\ref{f2:fig}.

Next, we consider the errors during the nonlinear regime of the
calculation. In particular, we consider the solution just prior to the
caustic formation, when the background expanded by a factor 25 since
the simulation start.  On the left hand side of Table~\ref{t3:tab} we report 
the $L_1$ and $L_\infty$ error norms and convergence rates
as in Table~\ref{t2:tab} while the $L-2$ errors are shown as open
symbols in the left hand side panels of Fig.~\ref{f2:fig}.
%
%
We see that in the nonlinear regime the convergence rates of the
errors in the particle positions, velocities and forces have worsened
in a minor, appreciable and considerable way, respectively.

Finally, we test the performance of the AMR code.  We use a constant
refinement ratio, $n_{\rm ref}=2$, and refine cells enclosing a mass
larger than 1.5 the average value.  A maximum of three levels of
refinement were allowed.  All runs used a first level of refinement
for about 30\% of the calculation, except for the lowest resolution
case for which the percentage was 12\%.  The second level of
refinement was only used by the three higher resolution runs and only
for about 5\% of the time.  Finally, a third level of refinement was
employed only by the two finest runs and only for less than 1\% of the
time. Similarly, finer grids cover a progressively smaller fraction
of the computational domain.
\begin{figure} [ht]
\begin{center}
\includegraphics[height=0.6\textheight, scale=1.0]{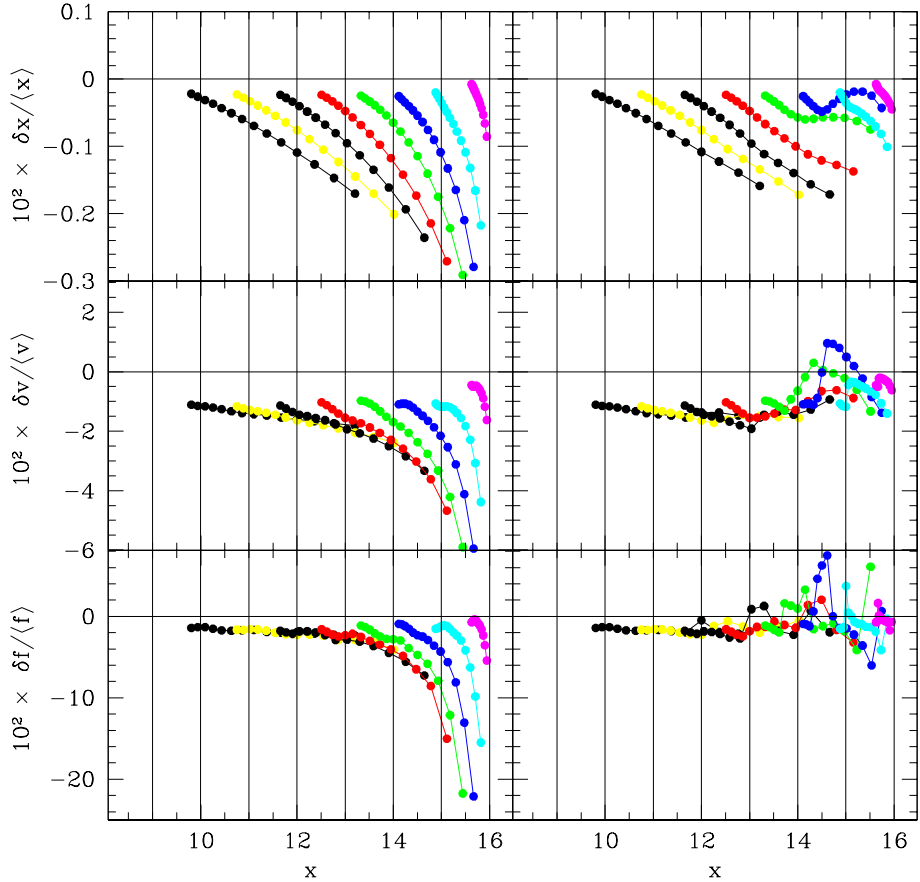}
\caption{Comparison between two numerical experiments, employing 
a uniform grid (left) and two levels of refinement (right), in terms
of error in the particle position (top), velocity (middle) and force 
(bottom). The initial set up includes 32 grid cells (bound by 
vertical lines) and 32 particles uniformly distributed. Only one 
quarter of the grid is shown, focusing on the critical region where 
the caustic forms. Vertical lines indicate cells' boundaries.
\label{f3:fig}
}
\end{center}
\end{figure}
The $L_1$ and $L_\infty$ error norms and convergence rates for the AMR
runs are reported on the right hand side of Table~\ref{t3:tab} while
the $L-2$ errors are shown as spur symbols in the left hand side
panels of Fig.~\ref{f2:fig}.  These results show that employing AMR
during the nonlinear evolution improves the convergence rate of the
solution in such a way that they resemble the values in the linear
stage. This is true in this example for the errors in the position
and velocity and to a lesser extent, the force, which is more affected
by coarse/fine boundary effects.

Fig.~\ref{f3:fig} compares the errors in position, velocity and
force for a fixed grid (left) and an AMR grid (right) calculations.  It
focuses on the region where, and the times when, the caustic forms and
AMR operates. Thus only one quarter of the total grid is shown (with
the cell boundaries of the base grid indicated by vertical lines and
marked by integer labels), and the various errors are plotted only
for the last one third of the simulation run. At the beginning of that
time span (when $a\sim 0.2$), the particles have clustered sufficiently
at the grid center and a first level of refinement is created.
%
\begin{figure} [ht]
\begin{center}
\includegraphics[height=0.6\textheight, scale=1.0]{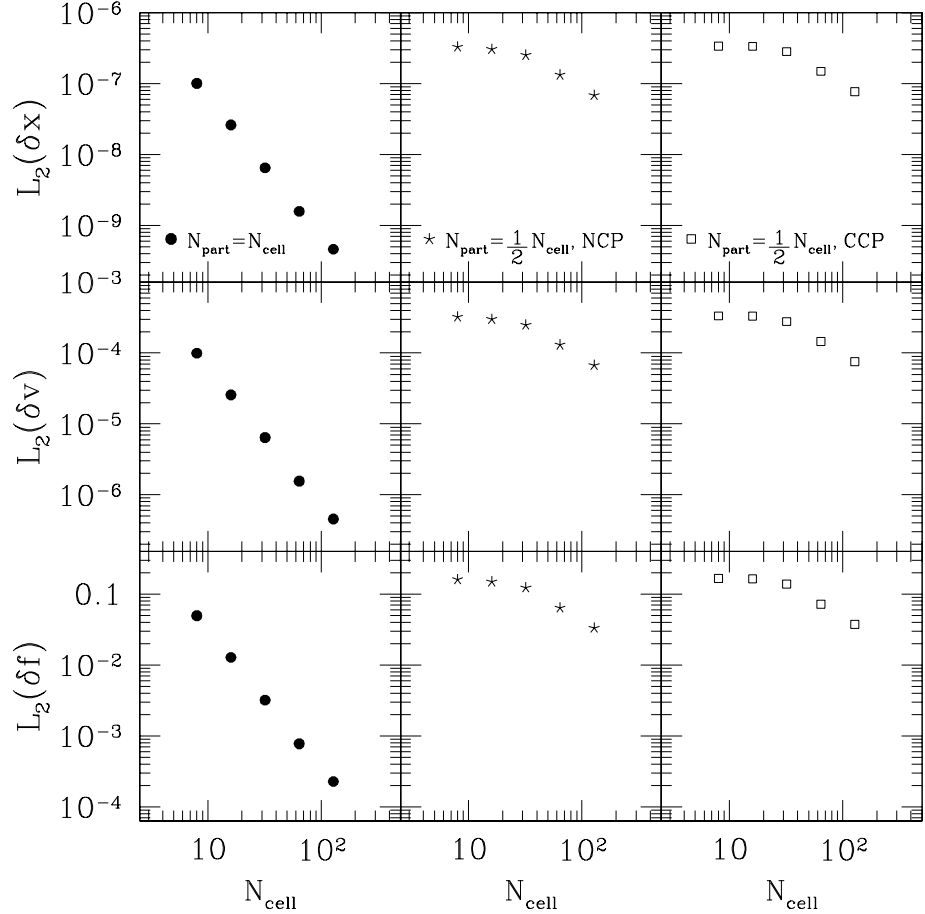}
\caption{
Same as Fig.~\ref{f1:fig} but for the case in which a 
staggered scheme is used for the force calculation.  
\label{f4:fig}}
\end{center}
\end{figure}
Six grid meshes are refined (only three between boundaries, 12-16, are
shown in Fig.~\ref{f3:fig}).  The second level of refinement is
created much later and only affects one base (or two refined) grid
mesh(es) for the last ten per cent of the simulation.
Fig.~\ref{f3:fig} shows that the errors in position, velocity and
force of the particles close to the point where the caustic forms are
much reduced when AMR is employed.
The generation of a level of refinement is accompanied by a change
in the mass distribution and the potential field. When this happens, a 
particle may experience a sudden change in terms of the force
field acting upon it.  These effects are responsible for the somewhat
`errant' behavior of the force error, as illustrated in the
bottom-left panel of Fig.~\ref{f3:fig}. Overall, though, for each
particle the force fluctuations in the AMR case are significantly
smaller than the force errors in the uniform grid case. 

As a last result for this section, in Fig.~\ref{f4:fig} we plot $L_2$
convergence errors analogous to Fig.~\ref{f1:fig} but for the case in
which the force was computed with a staggered scheme.  Comparison of
the two figures shows that when $N_{\rm part}=N_{\rm cell}$ the
solution obtained with the staggered scheme also converges with second
order accuracy, while being characterized by smaller errors.  This is
in agreement with previous findings~\cite{mellot86}.  However, when
the number of particles is halved ($N_{\rm part}=\frac{1}{2} N_{\rm
  cell}$) the results obtained with a staggered force scheme worsen
more dramatically than for the cell centered case, showing a very poor
convergence rate.  Closer inspection shows that when the particles are
sparse on the grid, oscillations appear in the potential due to the
discrete character of the matter distribution as reproduced on the
grid.  Since this affects only the quality of the staggered scheme the
problem may be related to the inconsistency of this scheme with the
centering of the stencil used for the discretization of the Laplacian
operator. We shall return to this issue in the next test case where
the problem reappears with more dramatic effects.

\subsubsection{Collisional Component}
\begin{figure} [ht]
\begin{center}
\includegraphics[height=0.6\textheight, scale=1.0]{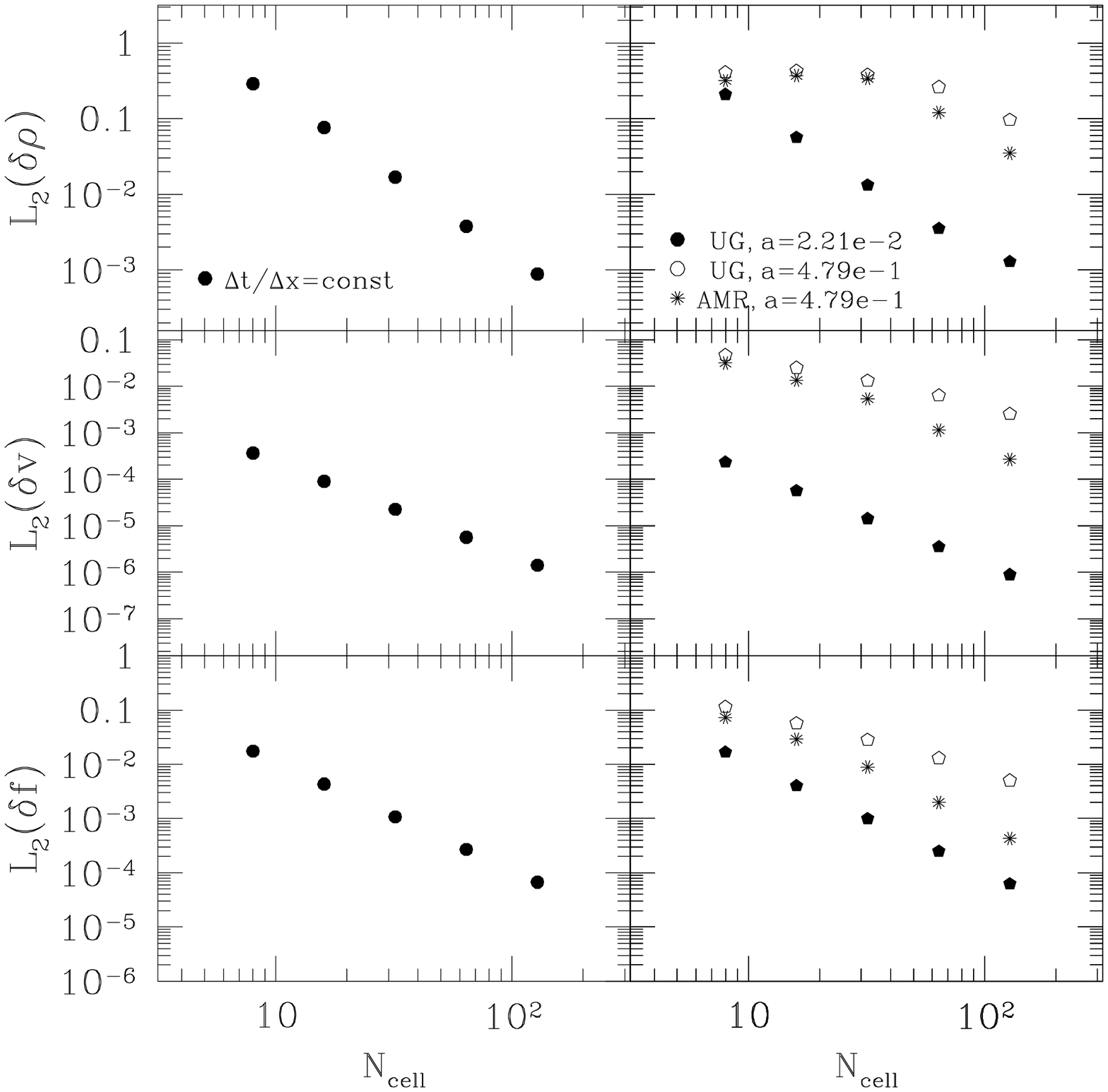}
\caption{
$L_2$ norm of the error in density (top), 
velocity (center) and force (bottom)
as a function of the number of grid cells.
See legend for the meaning of the symbols
(UG= Uniform Grid, $a$ is the expansion parameter).
\label{f5:fig}}
\end{center}
\end{figure}

We now turn to the performance of the hydrodynamic part of the code.
We use the same tests employed in the previous section for the the
collisionless component.  However, since in this case the solution
refers to an Eulerian grid, in order to have the velocity and density
at a given grid location $x(q,t)$ from Eq.
(\ref{vel.eq})-(\ref{rho.eq}) we must invert Eq. (\ref{pos.eq}).
\begin{table} [ht]
\label{god1:tab}
\caption{Convergence tests: Godunov's scheme$^\dagger$}
\begin{tabular*}{\textwidth}{@{\extracolsep{\fill}}lcccccc}
\hline 
\hline 
$N_{\rm part}$ & $L_1$ & $R_1$ & $L_2$ & $R_2$ & $L_\infty$ & $R_\infty$ \cr
\hline 
\multicolumn{7}{c}{\bf density}  \cr
8    &  2.7e-05 & 2.2 & 2.9e-05 & 1.9  & 4.2e-05 & 1.2   \cr
16   &  5.7e-06 & 2.2 & 7.6e-06 & 2.2  & 1.8e-05 & 1.8   \cr
32   &  1.2e-06 & 2.0 & 1.7e-06 & 2.2  & 5.3e-06 & 1.9   \cr
64   &  3.0e-07 & 2.0 & 3.8e-07 & 2.1  & 1.4e-06 & 1.9   \cr
128  &  7.4e-08 &  -- & 8.8e-08 &  --  & 3.7e-07 &  --   \cr
\hline
\multicolumn{7}{c}{\bf velocity}  \cr
8    &  3.3e-05 & 2.0 & 3.6e-05 & 2.0  & 5.2e-05 & 2.0   \cr
16   &  8.1e-06 & 2.0 & 9.0e-06 & 2.0  & 1.3e-05 & 2.0   \cr
32   &  2.0e-06 & 2.0 & 2.2e-06 & 2.0  & 3.2e-06 & 2.0   \cr
64   &  5.0e-07 & 2.0 & 5.6e-07 & 2.0  & 7.9e-07 & 2.0   \cr
128  &  1.2e-07 & --  & 1.4e-07 & --   & 2.0e-07 & --    \cr
\hline
\multicolumn{7}{c}{\bf force}  \cr
8    &  1.6e-02 & 2.0 & 1.7e-02  & 2.0  & 2.4e-02  & 2.0 \cr
16   &  3.9e-03 & 2.0 & 4.3e-03  & 2.0  & 6.1e-03  & 2.0 \cr
32   &  9.7e-04 & 2.0 & 1.1e-03  & 2.0  & 1.5e-03  & 2.0 \cr
64   &  2.4e-04 & 2.0 & 2.7e-04  & 2.0  & 3.8e-04  & 2.0 \cr
128  &  6.0e-05 & --  & 6.7e-05  & --   & 9.5e-05  & --  \cr
\hline
\hline 
\end{tabular*}
 \tablenotes{
 \qquad\llap{$^\dagger$} We use a cell-centered force scheme and a
constant timestep,$\frac{\Delta t}{\Delta x} = 1.6\times 10^{-4}$. 
 }
\end{table}
We first consider the errors in the linear regime, in 
analogy to Table~\ref{t1:tab} and Table~\ref{t2:tab} for the 
collisionless component.
In Table~\ref{god1:tab} we report the $L_1,~L_2$ and $L_\infty$ 
norms of the error for the gas density, velocity and force and the
corresponding convergence rates, for the case of fixed time step,
$\frac{\Delta t}{\Delta x} = 1.6\times 10^{-4}$.
Similarly, the left hand side of Table~\ref{god2:tab} reports the
$L_1,~L_2$ and $L_\infty$ errors for the case in which the time step
is set by the background expansion rate. The $L_2$ errors 
for the fixed and varying timestep cases are also
shown by the filled symbols in the left and right hand side panels of 
Fig.~\ref{f5:fig}, respectively.
These tests show the second order accuracy of the implemented scheme.
We note, however, that in the case of variable time step the 
convergence rate of the $L_\infty$ norm of the density error is slower.
\begin{table}  
\label{god2:tab}
\caption{Convergence tests: Godunov's scheme, linear regime$^\dagger$}
\begin{tabular*}{\textwidth}{@{\extracolsep{\fill}}lcccccc}
\hline 
\hline 
$N_{\rm part}$ & $L_1$ & $R_1$ & $L_2$ & $R_2$ & $L_\infty$ & $R_\infty$ \cr
\hline 
\multicolumn{7}{c}{\bf density}  \cr
8    &  2.0e-04 & 2.2 & 2.1e-04 & 1.9  & 3.0e-04 & 1.1   \cr
16   &  4.2e-05 & 2.1 & 5.7e-05 & 2.1  & 1.4e-04 & 1.7   \cr
32   &  9.5e-06 & 2.0 & 1.3e-05 & 1.9  & 4.3e-05 & 1.4   \cr
64   &  2.4e-06 & 1.9 & 3.5e-06 & 1.5  & 1.6e-05 & 0.9   \cr
128  &  6.5e-07 &  -- & 1.3e-06 &  --  & 8.4e-06 &  --   \cr
\hline
\multicolumn{7}{c}{\bf velocity}  \cr
8    &  2.1e-04 & 2.0 & 2.3e-04 & 2.0  & 3.3e-04 & 2.0   \cr
16   &  5.1e-05 & 2.0 & 5.7e-05 & 2.0  & 8.2e-05 & 2.0   \cr
32   &  1.3e-05 & 2.0 & 1.4e-05 & 2.0  & 2.0e-05 & 2.1   \cr
64   &  3.2e-06 & 2.0 & 3.5e-06 & 2.0  & 5.0e-06 & 1.9   \cr
128  &  7.9e-07 & --  & 8.9e-07 & --   & 1.3e-06 & --    \cr
\hline
\multicolumn{7}{c}{\bf force}  \cr
8    &  1.5e-02 & 2.0 & 1.7e-02  & 2.1  & 2.3e-02  & 2.0 \cr
16   &  3.6e-03 & 2.0 & 4.0e-03  & 2.0  & 5.7e-03  & 2.0 \cr
32   &  9.0e-04 & 2.0 & 1.0e-04  & 2.0  & 1.4e-03  & 2.0 \cr
64   &  2.2e-04 & 2.0 & 2.5e-04  & 2.0  & 3.5e-04  & 2.0 \cr
128  &  5.6e-05 & --  & 6.2e-05  & --   & 8.9e-05  & --  \cr
\hline
\hline 
\end{tabular*}
 \tablenotes{
 \qquad\llap{$^\dagger$} We use a cell-centered force scheme, a variable 
timestep, $\frac{\Delta t}{\Delta x} = C_{exp}(\frac{a}{\dot{a}}),
~C_{exp}=10^{-2}$, and $a=0.0221$.
 }
\end{table}
\begin{table}  [ht]
\label{godnl:tab}
\caption{Convergence tests: Godunov's scheme, nonlinear regime$^\dagger$}
\begin{tabular*}{\textwidth}{@{\extracolsep{\fill}}lccccccccc}
\hline
\hline
 & \multicolumn{4}{c}{uniform grid}  & \multicolumn{5}{c}{AMR} \cr
 \cline{2-5} \cline{6-10}
$N_{\rm part}$ & $L_1$ & $R_1$ & $L_\infty$ & $R_\infty$ 
 & $L_1$ & $R_1$ & $L_\infty$ & $R_\infty$ & $l_{max}$ \cr
\hline
 & \multicolumn{8}{c}{\bf density}  \cr
8    &  2.2e-01 & 0.5 & 8.2e-01 & --  &  1.6e-01 & 0.7 & 8.2e-01 & --  & 1\cr
16   &  1.6e-01 & 0.7 & 1.2e00  & --  &  1.0e-01 & 0.6 & 1.5e00 & --  & 1\cr
32   &  9.7e-02 & 0.9 & 1.5e00  & 0.0 &  6.4e-02 & 2.0 & 2.4e00 & 1.3 & 2\cr
64   &  5.0e-02 & 1.6 & 1.5e00  & 1.0 &  1.6e-02 & 2.0 & 1.0e00 & 1.1 & 2\cr
128  &  1.6e-02 & --  & 7.3e-01 & --  &  3.9e-03 & --  & 4.7e-01 & -- & 3\cr
\hline
 & \multicolumn{8}{c}{\bf velocity}  \cr
 8    &  2.7e-02 & 1.4 & 9.4e-02 & 0.4 &  1.6e-02 & 1.8 & 8.6e-02 & 0.7 & 1\cr
 16   &  1.0e-02 & 1.4 & 7.0e-02 & 0.4 &  4.7e-03 & 1.7 & 5.3e-02 & 0.6 & 1\cr
 32   &  3.8e-03 & 1.5 & 5.3e-02 & 0.6 &  1.4e-03 & 2.4 & 3.6e-02 & 2.2 & 2\cr
 64   &  1.3e-03 & 1.6 & 3.6e-02 & 0.9 &  2.6e-04 & 1.8 & 9.6e-03 & 1.9 & 2\cr
 128  &  4.3e-04 & --  & 2.0e-02 & --  &  7.2e-05 & --  & 3.1e-03 & --  & 3\cr
 \hline
 & \multicolumn{8}{c}{\bf force}  \cr
 8    &  6.8e-02 & 1.6 & 2.3e-01 & 0.5 & 3.7e-02 & 1.7 & 2.0e-01 & 0.9 & 1\cr
 16   &  2.3e-02 & 1.5 & 1.6e-01 & 0.5 & 1.1e-02 & 2.3 & 1.1e-01 & 0.8 & 1\cr
 32   &  8.4e-03 & 1.5 & 1.1e-01 & 0.6 & 2.2e-03 & 2.0 & 6.3e-02 & 1.7 & 2\cr
 64   &  2.9e-03 & 1.6 & 7.3e-02 & 0.9 & 5.5e-04 & 1.8 & 1.9e-02 & 2.2 & 2\cr
 128  &  9.5e-04 & --  & 3.9e-02 & --  & 1.6e-04 & --  & 4.0e-03 & -- & 3\cr
 \hline
\hline 
\end{tabular*}
 \tablenotes{
 \qquad\llap{$^\dagger$} We use a cell-centered force scheme, a variable
timestep, $\frac{\Delta t}{\Delta x} = C_{exp}(\frac{a}{\dot{a}}),
~C_{exp}=10^{-2}$, and $a=0.479$.
 }
\end{table}
%
%
%
Next, the left hand side of Table~\ref{godnl:tab} reports the errors
and convergence rates for the case of a uniform grid calculation,
well in the nonlinear regime, close the formation of the caustic
($a=0.479$).  As before, the $L_2$ errors are also shown as open symbols in
Fig.~\ref{f5:fig}.  Unlike the collisionless case, here is the
convergence rate of the density that is mostly affected, particularly
at low resolutions (cf.~\cite{bryanetal95,feshzh04}). 
As illustrated by the $L_\infty$ norm, the error is dominated
by the contribution of a few cells, located where the caustic forms.

Finally, we test the performance of the AMR code.  As for the
collisionless component, we use a constant refinement ratio, $n_{\rm
  ref}=2$, refine cells enclosing a mass larger than 1.5 the average
value and allow for a max of three levels of refinement.  The results
at $a=0.476$ are reported in the right hand side columns of
Table~\ref{godnl:tab} for the $L_1,~L_\infty$ errors and convergence
rates. There we also show the maximum level employed by each run.
$L_2$ errors are reported as spur symbols in the right panels of
Fig.~\ref{f5:fig}.

The use of refined grids in terms of fraction of grid covered and
fraction of the simulation time is very similar to the corresponding
collisionless case.  Similar is also the benefit of AMR, which
improves the convergence of the solution to rates very similar to
those characterizing the linear regime. This is indeed a powerful 
performance of the AMR technique.

\subsection{Effect of $C_{exp}$ on the Solution Quality}
We have investigated how the error depends on the choice of the
parameter $C_{exp}$ for the above problem. In particular we 
have computed the error
accumulated during an interval $\Delta a \ll a_{collapse}$, for values
of the expansion parameter $a=0.196$ and $a=0.091$, and for values of
$C_{exp}$ ranging from $10^{-3}$ to $0.5$. We consider both the fluid
and the collisionless case.
We use a uniform grid with 32 zones on a side and, for the collisionless
case, we use one particle per cell.
The results are rather independent of the norm type. 
We find that the errors introduced in
the particles velocity and position is quite stable, except for the
largest values of $C_{exp}$.  On the other hand, the errors in the
fluid components decrease steadily as $C_{exp}$ is reduced,
spanning a factor $\sim 5$ before reaching a plateau for $C_{exp}\leq
(1-2)\times 10^{-2}$.
\subsection{Homologous Dust Cloud Spherical Collapse}
In this section we test the ability of the code to follow the collapse of a 
pressure-less (dust) sphere of matter~\cite{cowh96}.
The problem is described by the following equations
\begin{eqnarray} \label{sc1.eq}
\left( \frac{\partial ^2 r }{\partial t^2} \right)_M  &  =  &  - G \frac{M(r)}{r^2} \\
u  &  =  &  \left( \frac{\partial r }{\partial t} \right)_M     \label{sc2.eq}
\end{eqnarray}
with initial conditions
\begin{eqnarray}  \label{scic1.eq}
\rho({\bf i}, t=0) &  =   & 
   \left\{ \begin{array}{lll} 
       f[r({\bf i})] & \mbox{if} &  r({\bf i}) \leq R \\ 
       0  & \mbox{if} &  r({\bf i}) > R
   \end{array} \right.   \\
   u({\bf i}, t=0)   &  =  &  0
\end{eqnarray}
where $M(r)$ is the mass enclosed within a distance $r$ from the
sphere center, $R$ is the radius of the sphere and $f(r)$ is a
function (with $f^\prime(r)\leq 0$) that depends solely on $r$.  For
the hydrodynamic case null density will be approximated with a value
$\rho(r>R) \ll \min[\rho(r)]$.  The problem admits a self similar
solution in implicit form which reads
\begin{eqnarray}
(1 - \xi ) ^{1/2} \, \xi^{1/2} + \sin^{-1} (1 - \xi ) ^{1/2} = \tau  \\
r=\xi \, r_0, ~~~u= \frac{r_0}{\tau_c} \,\left(\xi ^{-1} -1 \right) ^{1/2} , ~~~ \tau \equiv t \, \left( \frac{8\pi G \langle\rho\rangle_r}{3} \right)^{1/2} 
\end{eqnarray}
where $ \langle\rho\rangle_r= 3M(r)/4\pi r^3$ is the average density
within a radius $r$.

From the numerical point of view, the problem is challenging in two
respects: during collapse the force potential becomes progressively
steeper and, therefore, more demanding for the gravity solver. In
addition, since the problem has inherent radial symmetry and we are
solving it on a Cartesian grid, the ability of the code at preserving
that symmetry will be tested.
%
\begin{figure} 
\begin{center}
\includegraphics[width=1.0\textwidth,  scale=1.0]{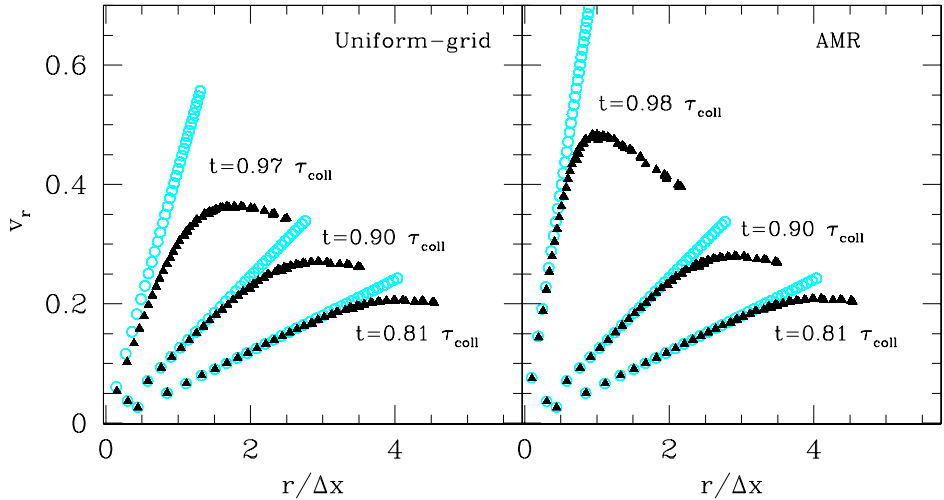}
\caption{ 
Phase space distribution of particles for the spherical collapse of a 
pressureless cloud. Cyan open circles correspond to the analytic 
solution and filled triangles to the numerical simulation result. 
The right panel is for the case of a uniform grid whereas 
the left panel correspond to an AMR calculation with two 
levels of refinement.
\label{dust_part:fig}}
\end{center}
\end{figure}

For the collisionless component we initially set
particles with null velocity at the center of cells whose distance
from the cloud center is less than $R$ (=1). This produces a
homogeneous density distribution everywhere inside R, except close to
the cloud edge due to the discreteness of the grid.  In 
Fig.~\ref{dust_part:fig} we compare the position of each particle in
phase-space ($v_r,r$) as given by the code (black filled triangles) 
with the analytic solution (cyan open circle).  The left panel corresponds
to the case in which a uniform grid is used whereas for the right panel
solution AMR was employed. We allow for two levels of refinement
and tag cells with a total mass four times as high as the initial value.
The comparison between analytic
and numerical solution in Fig.~\ref{dust_part:fig} is made for a number of
evolution times, expressed in terms of the adimensional collapse time
$\tau_{coll} = \pi/2$.  Noticeably, the code follows the particles
motions with high accuracy all the way down to the time of collapse.
In particular, there is no sign of artificial asymmetries.  Additional
levels of refinement were dynamically generated towards the final
phase of the collapse. At the latest time shown ($\tau = 0.98 \,
\tau_{coll}$), we can see the improvement due to the employment of
finer grids in the collapsing region.  Note that towards the edge of
the cloud the particles are trailing.  This is due to the inability to
reproduce a perfectly homogeneous sphere near the cloud edge from the
beginning.  The region affected by this is about one mesh size wide.

%
%
\begin{figure} 
\begin{center}
 \includegraphics[width=1.0\textwidth,  scale=1.0]{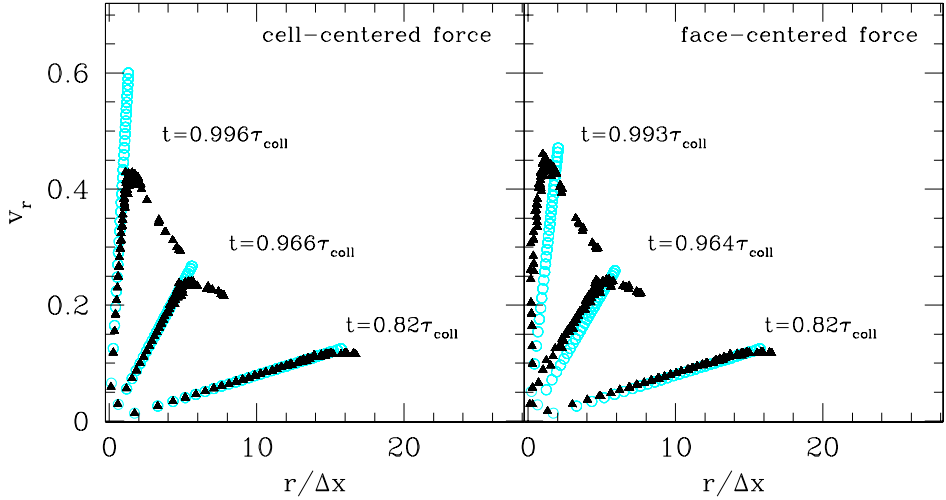}
\caption{
Phase space distribution of particles for the spherical collapse of 
a pressureless cloud. The initial cell-to-particle ratio is 4. 
Cyan open circles correspond to the analytic solution
and filled triangles to the numerical simulation result using
a cell-centered (left) and face-centered (right) force scheme, respectively.
\label{ccvsss:fig}}
\end{center}
\end{figure}

When exploring the accuracy of cell-centered versus face-centered force
schemes our tests suggest that, again, in the uniform grid case the latter perform 
slightly better, at the level of ca 15\%.
In analogy with the analysis of Sec.~\ref{zeld:sec}, 
we have tested this further, for the case in which the ratio of cells to
particles is significantly larger than one. This situation may easily
occur depending on the adopted criterion for refinement and on the
efficiency for the generation of the refined grid out of the tagged
cells. In Fig.~\ref{ccvsss:fig} we compare the solutions obtained with a
cell-centered (left) and a staggered (right) force scheme for an
initial cell-to-particle ratio of 4. The initial dust sphere is placed
on a grid of 256 cells on a side, and 8 particles are aligned along
its radius out to 32 cells from its center.  The code output is
plotted at three different solution times close to the time of
collapse $t=\tau_{coll}$.  This test shows that when the number of
cell-to-particle ratio is significantly higher than one, the staggered
force scheme tends to produce spurious results.  This is in line with
the findings in the previous test problem in section~\ref{zeld:sec}.
On the other hand, the cell-centered scheme seems well behaved.  
We find that the qualitative result does not change when we use a
two or four point force stencil, when the number of particles to resolve the
sphere is changed or when the sphere center is shifted by a fraction
of a mesh size in an arbitrary direction.

Thus the staggered force scheme, although apparently more accurate
than its cell-centered counterpart when the number of cells is
comparable to, or less than, the number of particles~\cite{mellot86}, 
it gives rise to spurious results when particles are sparse on the grid.
Therefore, caution must be exercised when employing staggered 
schemes for force evaluation.

%
%
\begin{figure} [ht]
\begin{center}
 \includegraphics[width=1.0\textwidth,  scale=1.0]{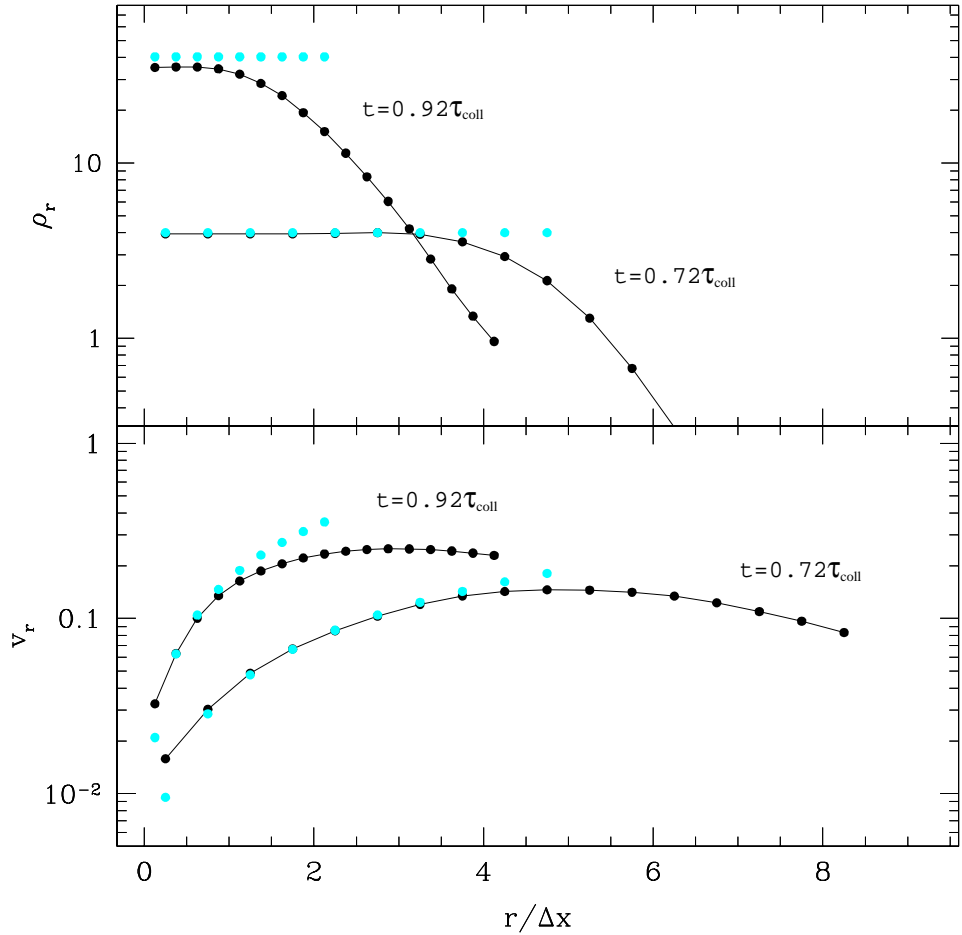}
\caption{
Density and velocity profile for spherical collapse of a pressureless 
cloud of gas. Cyan open circles correspond to the analytic solution and 
black filled dots to the numerical simulation results.
\label{dust_gas:fig}}
\end{center}
\end{figure}
Finally, the results for the collisional case are illustrated in
Fig.~\ref{dust_gas:fig}.  The plot compares the density (top) and
velocity (bottom) profiles of the numerical solution (threaded black
dots) and the analytic solution (cyan dots). The latter extend only
out the cloud size, whereas the numerical solution includes the region
covered by the finest level.  Two times during the collapse are shown:
$t=0.72\tau_{coll}$ and $t=0.92\tau_{coll}$, corresponding to the low
and high curves, respectively.  At these times one and two levels of
refinement have been generated, respectively. The chosen times are
close to the collapse time, when the errors have accumulated and the
simulation becomes more challenging.  Nevertheless, as for the
collisionless component, the code follows accurately the evolution of
the density and velocity profiles of the collapsing cloud. Again,
close to the cloud edge the density profile is smoother and the
velocity field slower than the analytic solution. The size of the
region affected by this is again of order of the coarse mesh size and
it is partially ascribed to the crude representation of the cloud edge
on the grid.

\subsection{Energy Conservation: Layzer-Irvine Equation}
For pure hydrodynamics conservation of the total (kinetic+thermal)
energy is enforced by our conservative Godunov's method.
When gravity is added, energy conservation should still hold, but
is not explicitly enforced in our scheme. Finally, 
with an expanding background energy is not conserved. 
For a collection of particles that interact only gravitationally,
the evolution of the energy of the system is regulated by the 
Layzer-Irvine equation, which reads
\begin{equation} \label{li:eq}
\frac{d}{dt} [a(t) ({\cal E + W}) ] = -\dot{a}\, {\cal E}
\end{equation}
where ${\cal E}$ is the kinetic energy associated to the peculiar
motions of the particles and ${\cal W}$ their gravitational potential
energy due to the overdensity produced by their mass distribution.  
Clearly in absence of expansion ($a=1$, $\dot{a}=0$)
Eq.~(\ref{li:eq}) reduces to the ordinary energy conservation
equation.  Otherwise it describes the change in the total energy of
the system due to the adiabatic expansion of the background.  The
derivation and physical meaning of Eq.~(\ref{li:eq}) is reviewed in
Ref.~\cite{peebles93}. Its applicability to hydrodynamic simulations
is discussed in, e.g., Ref.~\cite{rokc93}, in which case a monoatomic
gas is assumed and ${\cal E}$ includes both the kinetic and thermal
energy of the gas.

Eq.~(\ref{li:eq}) can be integrated in time giving
\begin{equation} \label{li_int:eq}
a [{\cal E}(a) + {\cal W}(a)] -a_0 [{\cal E}(a_0) + {\cal W}(a_0)] 
= - \int_{a_0}^{a} {\cal E} da.
\end{equation}
We can evaluate the integral on the RHS of the above equation
numerically with the trapezoidal rule and assess the accuracy of the
code at tracking the energy of the system through the quantity
\begin{equation} \label{li_err:eq}
\delta \varepsilon = 
\frac{a [{\cal E}(a) + {\cal W}(a)] -a_0 [{\cal E}(a_0) + {\cal W}(a_0)]+\int_{a_0}^{a} {\cal E} da}{[a_0 {\cal W}(a_0) - a {\cal W}(a)]} .
\end{equation}
%
%
\begin{figure} 
\begin{center}
 \includegraphics[width=0.5\textwidth,  scale=1.0]{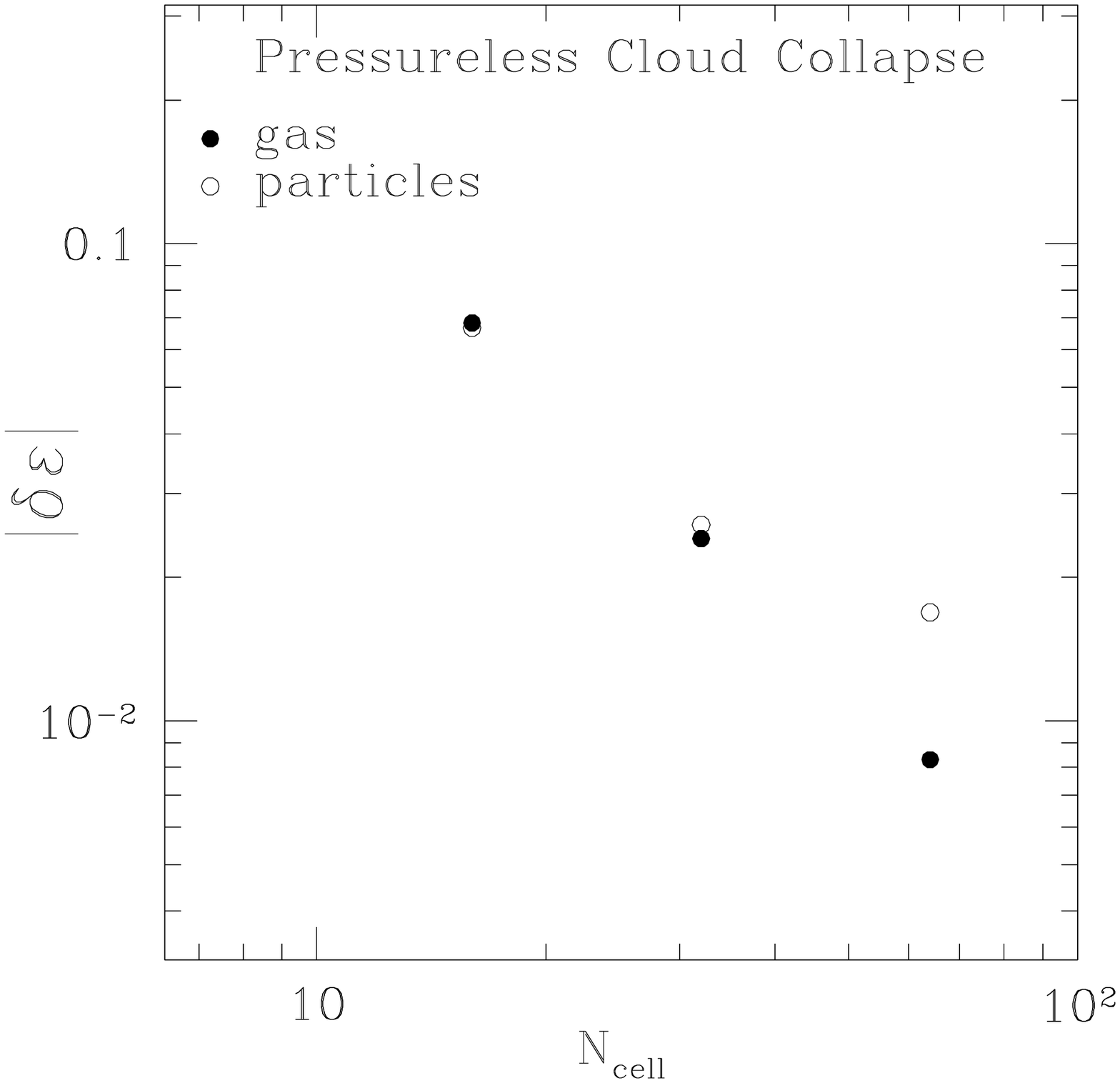}\includegraphics[width=0.5\textwidth,  scale=1.0]{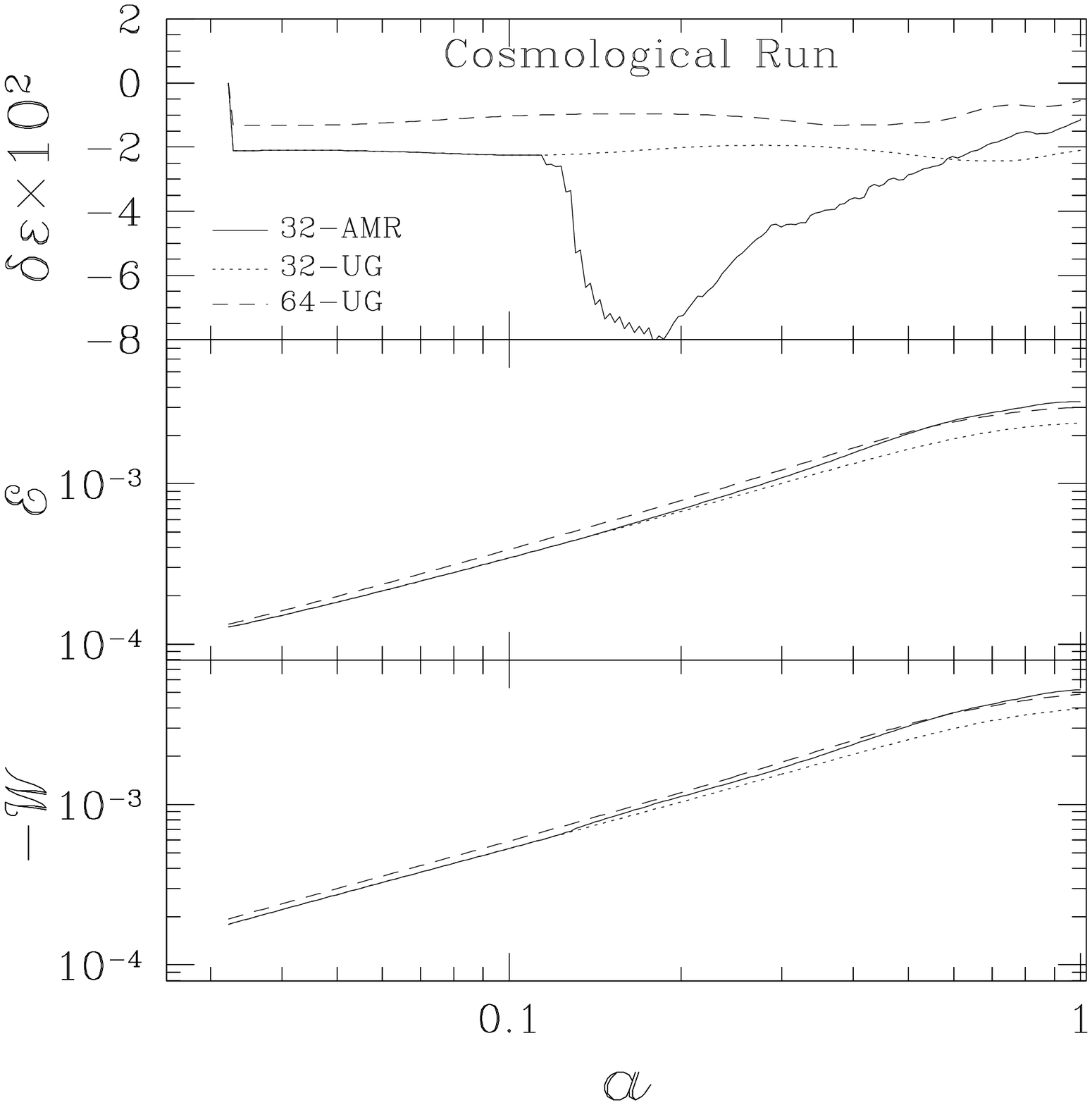}
\caption{Energy conservation error based on Eq.(\ref{li_err:eq})
for the collapse of a pressureless spherical cloud (left) and a cosmological
run (right). See text for details. \label{energy_error:fig}}
\end{center}
\end{figure}

We first test the energy conservation accuracy of the code for the
case of the collapse of a pressureless cloud.  This is the problem
studied in the previous section.  We carry out three AMR
calculation with different base grid sizes, namely 16, 32, 64
corresponding to 4, 8, 16 cells per cloud radius respectively. In
these runs, cells enclosing more then four times the initial mass
content were tagged for refinement and a maximum of two 2 refinement
levels were allowed.  The results of the test are reported in the left
hand side panel of Fig.~\ref{energy_error:fig} where we plot the error in the
total energy, $\delta \varepsilon$, as a function of resolution, for
both the particle (open) and the gasdynamic (filled) case.  The plots
show that with 16 zones per cloud radius the error in the
energy is at the level of a per cent or so.

Next we test code accuracy at tracking the energy of the system
in a cosmological run. For the
purpose we use a $\Lambda$-Cold Dark Matter cosmology with parameters
$\Omega_m$=0.3,~$\Omega_\Lambda=0.7$,~$\Omega_b$ = 0.04, for the
energy density in total matter, dark energy and baryonic matter
respectively; and $H_0 = 70$ km s$^{-1}$ Mpc$^{-1}$ for the Hubble
constant.  The physical domain has a size of L=91.43 Mpc on a side.
We execute three runs with different numerical resolution.  The first
two runs employ a uniform grid with 32$^3$ and 64$^3$ cells,
respectively, and the same number of particles as grid cells.  The
third run uses a base grid with 32$^3$ cells and 32$^3$ particles, and
two additional levels of refinement created in region where the total
mass enclosed in a cell exceeds the initial value by a factor eigth.
The initial conditions where generated on a 64$^3$ grid and
coarse-averaged to a 32$^3$ grid for the low-resolution-uniform and
AMR runs.

The results are presented in the right panel of
Fig.~\ref{energy_error:fig} where we plot $\delta \varepsilon$ (top),
${\cal E}$ (middle) and ${\cal W}$ (bottom) as a function of expansion
parameter $a$ for each resolution case. The plots show that when using
a uniform grid the total energy of the system is evolved with an
accuracy at the percent level ($\sim$ 2\% and 1\% for the $32^3$ (dot)
and $64^3$ (dash) cases, respectively), with most of the error
generated at startup.  In the AMR case (solid line), however, our
error parameter $\delta\varepsilon$ increases visibly when refinement
levels are created (at $a\sim 0.15$ and $a\sim 0.2$ for the first and
second level respectively).  The reason for this is simple. When a
level of refinement is created the potential energy of the system
changes suddenly (see solid and dot lines in the bottom panel)
throwing off the balance bewtween particle/gas velocities and their
potential energy. As a result a large error in the sense of
Eq.(\ref{li_err:eq}) is generated. This is so, even though with the
additional level of refinement the potential energy of the system is
more accurate as it gets closer the value from the high resolution run
(see solid and dash lines in the bottom panel). Over time the kinetic
energy readjusts to the new potential (middle panel; notice that the
internal energy is negligible) and a balance between the two forms of
energy is reestablished.  Because the potential energy associated with
the newly formed structures is larger than that of the system at the
time when the refinement levels were first generated, the new balance
between kinetic and potential energy reduces substantially the error
as time progresses. At simulation end the AMR run (with a base grid of
$32^3$ cells) produces estimates of $\delta \varepsilon$, ${\cal E}$
and ${\cal W}$ very close to the high resolution run.

\subsection{Santa Barbara Galaxy Cluster}
In this section we carry out the calculation defined by the
`Santa Barbara Cluster Comparison Project'~\cite{frenketal99}
and compare the results of our code with those from different
codes implemented independently by other authors 
and based either on similar or different techniques.

The problem consists of simulating the formation of a galaxy cluster
in a Standard Cold Dark Matter universe.  The cosmological parameters
assumed were $\Omega_m$= 1 and $\Omega_b$ = 0.1 for the total and
baryonic mean mass density in units of the critical density,
respectively; $H_0 = 50$ km s$^{-1}$ Mpc$^{-1}$ for the Hubble
constant; $\sigma_8$ = 0.9 for the present-day linear rms mass
fluctuation in spherical top hat spheres of radius 16 Mpc; for the
baryon density.  The computational domain has a size of L=64 Mpc on a
side.  The initial matter fluctuation are characterized by a power
spectrum with an asymptotic spectral index, n = 1, and are
`constrained' so that at simulation end a massive structure has formed
at the center of the computational box.

The simulation was initialized at z = 40 with two grids already in
place: a base grid covering the entire 64 Mpc$^3$ domain with 64$^3$
cells and 64$^3$ particles; and a second grid, also with 64$^3$ cells
and 64$^3$ particles, but only 32 Mpc on a side and placed in the
central region of the base grid, thus yielding an initial cell size of
0.5 Mpc.  Refinement is applied only in this central, higher
resolution region and is based on a local density criterion: cells
with a total mass of $6.4\times 10^{10}$ M$_\odot$ or more were
refined.  We allowed for 5 levels of refinement (for a
total of a 6 levels hierarchy), with a constant refinement ratio
$n_{ref}=2$.  The size of the finest mesh is about 15 comoving
kpc.  We use the following CFL coefficients for the time step: $C_{\rm
  hydro}=0.8,~C_{\rm part} =0.5$ and $C_{\rm exp}=0.02$.

\begin{figure} [ht]
\begin{center}
\includegraphics[height=1.\textwidth, scale=1.0]{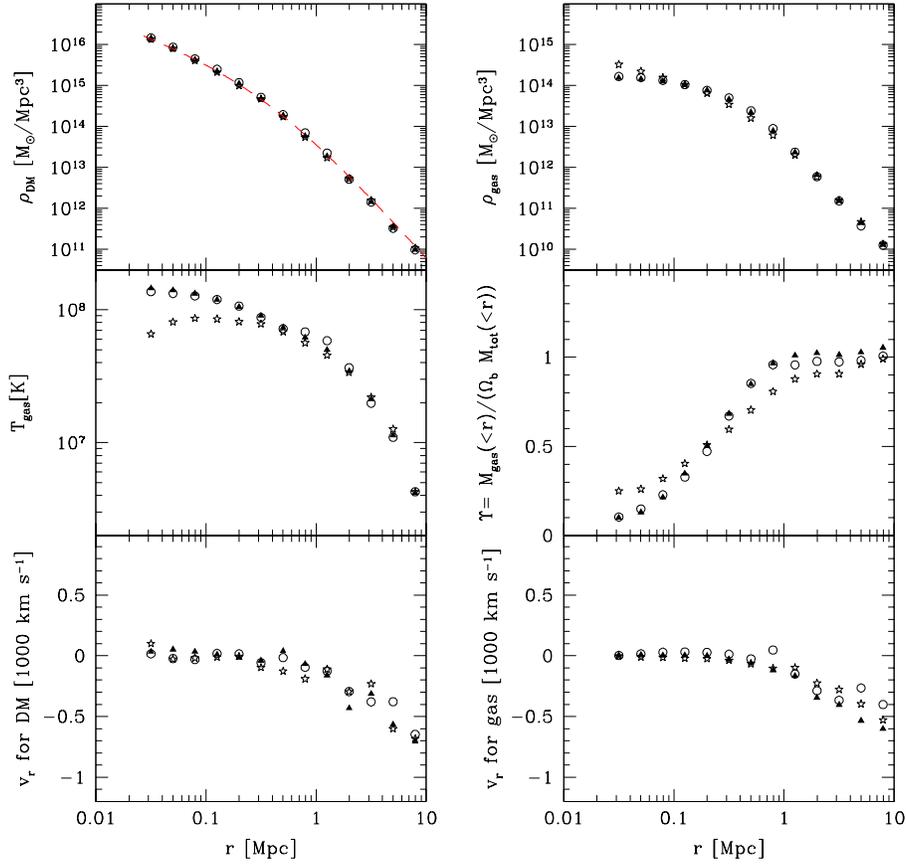}
\caption{Radial profile of dark matter (top left), baryonic gas (top right)
temperature (middle left), baryonic fraction (middle right), radial velocity
for dark matter (bottom left) and gas (bottom right). 
In addition to the results from {\tt CHARM} (open dots), 
for comparison we also show those from the {\tt ENZO} 
AMR code (filled triangles)~\cite{bryannorman97} 
as well as those from the {\tt HYDRA} SPH code (open stars)~\cite{cope95}. 
\label{sbplots.fig}}
\end{center}
\end{figure}

At simulation end a halo finder based on the spherical overdensity
method~\cite{laco94} was run in order to define the center of the
galaxy cluster.  The radial profiles for six quantities of interest are
presented in Fig.~\ref{sbplots.fig} together with results 
from two other simulation codes: {\tt ENZO}, which is an Eulerian
AMR code similar to ours, and {\tt HYDRA} (as run by Jenkins \&
Pearce), which combines smoothed particle hydrodynamics (SPH) and
adaptive particle-particle-particle-mesh (AP$^3$M) method for the
N-body part~\cite{cope95}.  These two codes are meant to be
representative of the high resolution grid based and SPH approaches,
respectively.
%
%

Results are shown down to scales of about 30 kpc which is just above
the nominal resolution at finest level of refinement.  The plot shows
that there is good agreement among the results of the different codes
particularly with {\tt ENZO} (even though we used one less level of
refinement and use a different refinement criterion).  The discrepancies,
when significant, are consistent with those already found in
the extensive comparison paper in Ref.~\cite{frenketal99}.  In fact,
there is very good agreement in terms of dark matter density
distribution (top left) which is well fit by the analytical form
proposed in Ref.~\cite{nfw97} with parameters specified in the
caption. Similarly, there is good matching of the solution in terms of
gas density distribution, except in the inner regions within 100 kpc,
where the two AMR solutions flattens and the SPH solution keeps
on increasing. More significant is the difference in temperature
distribution (middle left), which drops in the SPH solution for the
inner regions and stays constant for the AMR case.  These differences
were already found and discussed in Ref.~\cite{frenketal99}.  (See
also Ref.~\cite{quilis04} for similar findings.) Their origin is
not fully clear, although as suggested in Ref.~\cite{frenketal99}, it
could be ascribed to the different way in which SPH and grid based
methods treat shocks. 

Next panel (middle right) shows the profile for the ratio of gas to
dark matter mass, normalized to the global value. While our solution
is in good agreement with {\tt ENZO}'s and deviates from {\tt HYDRA}'s
in the inner regions and is somewhat in between the two beyond 1 Mpc
or so.  There is significant scatter in the results from the
full set of codes found in Ref.~\cite{frenketal99}, at the level of
0.1-0.2. Nevertheless, it is pointed out in Ref.~\cite{frenketal99}
that when integrated out to the virial radius of the system (2.7 Mpc), 
the SPH codes seem to predict a systematically slightly smaller values 
for this quantity than high resolution grid based codes. The reason for
this is still not clear.

Finally, both the gas and dark matter radial velocity profiles agree 
quite well at all radii. Some differences may arise due to slight 
differences in the simulation timing, as pointed out in 
Ref.~\cite{frenketal99}. Note that at larger radii (last few points), 
typically characterized by wider scatter, both ENZO's and HYDRA's 
results tend to be below the average value defined in
Ref.~\cite{frenketal99}.

\section{Conclusions}

We have presented a new code based on AMR technique for systems
comprising collisional and collisionless components coupled through a
long range force. We have thus extended the scheme
in~\cite{bergercolella89} to include collisionless particle dynamics
and gravity arising from the mass distribution of the two components.
For the hydrodynamics we use a slightly modified directionally unsplit
Godunov's method based on Ref.~\cite{colella90}.  As for the
collisionless component we have implemented various time centered
modified symplectic schemes based on both the kick-drift-kick and
drift-kick-drift sequence. Our implementations of these schemes appear
to perform comparably.  We have also used several types of stencil to
calculate the force from the potential.  We find that while the
staggered schemes appear more accurate when the number (density) of
particles is at least as large as the number of grid cells, it
produces spurious results when the particles are sparse on the grid.
Cell centered stencils thus seem more reliable, especially when a five
point cell centered discretization of the Laplacian operator is used.

Due to the time refinement character of the AMR technique the solution
on different levels is advanced with different timesteps.
Synchronization issues then arise as the multilevel solution to the
elliptic equation needs to be solved simultaneously on all levels.  In
particular, the density field represented by the particles evolved on
finer levels may not be available on coarser levels when they are not
synchronized.  Similarly, one cannot account for the effects of the
mass distribution on finer levels on the multilevel solution of the
potential, unless all levels are synchronized.  Among other features
of the code, we have thus introduced an {\it aggregation} procedure to
cost-effectively represent on the coarser levels the particles on
finer levels without compromising the code accuracy and performance.
We have also implemented a procedure for estimating (when the coarse
and finer levels are not synchronized) the effects on the coarse
potential produced by the matching conditions at refinement boundaries
between coarse and fine solution, as they would arise in a full
multilevel calculation.  We performed several standard tests which
illustrate the code accuracy as well as the advantages of the AMR
technique for the study of both self gravitating hyperbolic systems,
collisionless system and hybrid systems.







\begin{acknowledgment}
FM is thankful to D. Serafini, D. Martin, B. van Straalen and
D. Graves for useful discussions,  to the Lawrence
Berkeley National Laboratory, 
for its hospitality, to the Institute of Informatics, ETH Z\"urich, 
for technical support.
FM also acknowledges partial support by the
European Community through contract HPRN-CT2000-00126 RG29185
and by the Swiss Institute of Technology through a Zwicky
Prize Fellowship.
\end{acknowledgment}

\appendix{Charge Assignment Schemes} \label{cas.ap}

Indicating with $\Delta x$ the mesh size
in one dimension we find
the nearest grid point scheme (NGP) defined for order $r=1$ as 
\begin{equation}
 W(\hat x - x) = 
\left\{ \begin{array}{ll} 
1&  ~~~~~~~~~ {\rm if}~~  |\hat x -x| \leq \Delta x/2 \\
0&  ~~~~~~~~~ {\rm otherwise}  
\end{array} 
\right.
\end{equation}
in which the  assigned charge distribution 
is discontinuous as the particle cross the cell boundary;
the cloud in cell (CIC) scheme defined for $r=2$ as 
\begin{equation}
 W(\hat x - x) = 
\left\{ \begin{array}{ll} 
1-\frac{|\hat x -x|}{\Delta x} ~~ {\rm if}~~ & |\hat x -x| \leq \Delta x \\
0&  ~~ {\rm otherwise}  
\end{array} 
\right.
\end{equation}
in which the assigned charge distribution 
is continuous but the first derivative is not; 
the cell boundary;
and the triangular shape cloud (TSC) scheme defined for $r=3$ as 
\begin{equation}
 W(\hat x - x) = 
\left\{ \begin{array}{ll} 
\frac{3}{4}-\left(\frac{|\hat x -x|}{\Delta x}\right)^2 
&  ~~~~~~~~~ {\rm if}~~  |\hat x -x| \leq \Delta x/2 \\
\frac{1}{2}-\left(\frac{3}{2}-\frac{|\hat x -x|}{\Delta x}\right)^2 
&  ~~~~~~~~~ {\rm if}~~  \Delta x/2 \leq  |\hat x -x| \leq 3\Delta x/2 \\
0&  ~~~~~~~~~ {\rm otherwise}  
\end{array} 
\right.
\end{equation}
in which both the assigned charge distribution 
and first derivative are continuous. These schemes retain their 
properties when they are extended to a multidimensional case in the 
form of a product
\begin{equation}
W({\bf \hat x} - {\bf x}) = \prod_{j=1}^D W_j(\hat x_j - x_j).
\end{equation}

%
%

\bibliography{../biblio/papers,../biblio/books,../biblio/proceed,../biblio/codes}

\end{article}
\end{document}